\documentclass[12pt]{article}
\usepackage[utf8]{inputenc}
\usepackage{booktabs}
\usepackage{makecell}
\usepackage{tabularx}
\usepackage[font=small,labelfont=bf]{caption}
\usepackage{tikzfill.image}
\usepackage{xcolor}
\usepackage[most]{tcolorbox}
\usepackage{fancyvrb}
\usepackage{amssymb,amsmath,amsfonts,eurosym,geometry,ulem,graphicx,caption,color,setspace,sectsty,comment,footmisc,caption,natbib,pdflscape,subfigure,array, booktabs, graphicx}
\usepackage[hidelinks]{hyperref}

\tcbset{
  colback=gray!10,    
  colframe=black,     
  boxrule=0.8pt,      
  arc=4pt,            
  left=6pt,right=6pt,top=6pt,bottom=6pt
}

\newcolumntype{L}[1]{>{\raggedright\let\newline\\arraybackslash\hspace{0pt}}m{#1}}
\newcolumntype{C}[1]{>{\centering\let\newline\\arraybackslash\hspace{0pt}}m{#1}}
\newcolumntype{R}[1]{>{\raggedleft\let\newline\\arraybackslash\hspace{0pt}}m{#1}}

\begin{document}

\begin{titlepage}
\title{AI in the Enterprise:\\How People Use M365 Copilot Chat\thanks{We thank David Tittsworth for classification pipeline engineering and Dhruv Joshi for language on data privacy and general feedback.}}
\author{Scott Counts, Yan Chen$^{\dagger}$, Jing Dong$^{\dagger}$, Himanshu Sharma$^{\dagger}$\\ Andrey Zaikin$^{\dagger}$, Rui Hu, Alperen Kok, Gorkem Ozer Yilmaz,\\Siddharth Suri, Kiran Tomlinson, Sonia Jaffe, Will Wang
\\
\footnotesize{$^{\dagger}$Equal contribution, alphabetical by last name}\\
\normalsize{Microsoft Corporation}\\
\normalsize{One Microsoft Way, Redmond, WA 98052, USA}
}

\date{} 
\maketitle
\begin{abstract}

\noindent M365 Copilot is used every week by millions of people across more than a million companies around the world as part of their workflows. Uniquely positioned in the AI landscape given its near-exclusive use for work purposes, M365 Copilot can offer a clear picture of how people use AI for work and where that usage may expand next. This paper characterizes that usage through direct classification of user interactions with M365 Copilot Chat. Based on an anonymized and privacy-preserving analysis of a sample of approximately 5.5 million sessions, we combine a learned classification of user intent with a classification of O*NET work activities done with M365 Copilot Chat. We find that M365 Copilot is emerging as an everyday assistant for knowledge work: writing dominates, but users also rely on it for information retrieval, analysis, decision making and strategizing, and evaluating and diagnosing programs and systems, among others. Information seeking tasks remain common, but time trends suggest a relative shift away from ``chat as search'' and toward content and communication-related work. Comparisons across occupational groupings and to work done in the labor market further show that usage is broad but uneven, where the relative share of work done with M365 Copilot Chat cuts across jobs in some cases and is occupation-specific in others. Areas of relative underrepresentation in the labor market suggest the next frontier for enterprise AI adoption.
\end{abstract}

\setcounter{page}{0}
\thispagestyle{empty}
\end{titlepage}
\pagebreak \newpage


\section{Introduction} \label{sec:introduction}
M365 Copilot is an AI platform used by people in over a million companies around the world as part of their weekly workflows. As such, its usage offers a distinctive window into one of the most important questions of our time: how is AI being incorporated into everyday work? To answer that question, we need direct measurement of real-world enterprise AI usage to assess what people are using AI for, how those uses cluster across kinds of work, and where adoption remains thin relative to the work people do. Thus, our aims are two-fold: first, to present an empirical characterization of M365 Copilot Chat\footnote{Referred to throughout for simplicity as M365 Copilot.} based on direct measurement of user-Copilot conversations; and second, to use that characterization to identify where enterprise AI is already becoming an everyday assistant for knowledge work and where deeper adoption may still lie ahead.

These questions are important because the impact of AI on work may be profound. The growth of M365 Copilot, along with other AI services, has pushed AI to the forefront candidate as the next general-purpose technology \citep{bresnahan1995general, eloundou2024gpts}. Estimates of the resulting economic impact range from small but non-trivial \citep{acemoglumacroai} to substantial \citep{collisaicontribution} to the possibility of economically transformative productivity growth at 3 -- 5x historical averages \citep{brynjolfssonagendaai}. These estimates vary widely and are likely to shift as frontier capabilities emerge, as cost for training and inference at scale decreases, and as the technology becomes embedded in workflows. What remains especially important, then, is understanding not only what AI may be capable of in principle, but how it is already being used in practice.

AI has already seen considerable adoption, both generally \citep{chatterji2025chatgpt} and for work specifically. Results from the Real-Time Population Survey show that 23\% of workers in the U.S. used AI weekly for work in 2024, with 9\% using it daily \citep{bick2024rapid}. Similarly, a Gallup poll found that the percent of U.S. employees that use AI at work a few times per year or more has doubled since 2023 to 40\% in 2025, with 19\% using it weekly and 8\% daily \citep{gallup2025}. Further, emerging research shows this usage translates to productivity gains and time savings \citep{dillon2025early}, both generally \citep{jaffegenaiworkplaces} and in specific task areas like writing \citep{noy2023experimental} and domains such as software development \citep{peng2023impact, cui2024effects} and customer support \citep{brynjolfsson2025generative}. But usage data from broad consumer platforms can only partially answer how AI is being used inside work itself. M365 Copilot is unusually informative in this regard because its usage is overwhelmingly work-related and embedded in enterprise workflows.

Our analyses show that M365 Copilot is becoming an everyday assistant for knowledge work. Across user intents and O*NET work activities, the most common uses center on writing and editing, information retrieval including company-specific information seeking, analysis, decision making and strategizing, and communication. When we distinguish between the user side of the conversation (user goal) and the M365 Copilot side (Copilot action), the activities that most strongly skew toward Copilot center on providing information, advice, and other forms of assistance. Taken together, these patterns suggest that enterprise AI is already being used not as a narrow capability for a small set of tasks, but as a broadly useful layer in day-to-day knowledge work.

At the same time, our data show that usage is broad but uneven. Prior work already suggests that AI use is concentrated in knowledge work and communication tasks and is far less prevalent in physical-world work \citep{tomlinsonworking,handa2025economic,eloundou2024gpts}. Our findings corroborate this skew, but they also sharpen it by showing where enterprise AI appears underused relative to the work done in industries, occupations, and the labor market more broadly. Comparing M365 Copilot activity with work distributions in Banking, Consulting, and Manufacturing, and across occupational groupings, reveals both work activities that are absent from current usage and activities that are present but underrepresented. These underrepresented areas point to a next frontier for enterprise AI: not adoption in general, but deeper adoption across underused work.

Most companies are comprised of many job roles regardless of industry. We thus complement our industry comparison by looking at usage across occupational aggregations. Here we find usage patterns that both cut across and are uniquely indicative of different job families, again highlighting the broad \textit{and} specialized nature of M365 Copilot usage and also suggesting areas for job-specific AI functionality. We compare the share of M365 Copilot usage in different work activities to expectation based on the fraction of the labor force that do those work activities. Some activities such as those related to \textit{Getting Information} are overrepresented, while others fall below expectation. Work activities underrepresented compared to the size of the relevant workforce present a third view on areas for potential growth of AI.

Our ability to answer these questions around the impact of AI on work is enabled by direct measurement of the activities done with AI for work.
We do this through two classification schemes, O*NET \citep{onet29} from the U.S. Department of Labor and a bottom-up learned intent taxonomy to directly measure the types of tasks done. The O*NET taxonomy allows us to compare the amount of different types of work done with M365 Copilot to the amount of the same work in the labor market as a whole (see \citet{brynjolfsson2017ml,brynjolfsson2018machines} for early use of this technique), and to use of other AI platforms characterized with the same classification taxonomy \citep{chatterji2025chatgpt,handa2025economic,appel2025anthropic,anthropic2026aeiv5,anthropic2026aeiv4,tomlinsonworking,eloundou2024gpts}. The intent taxonomy provides a general characterization of the nature of usage---what are people doing and how frequently---that supports a broad-based understanding in a taxonomy native to AI usage in the unique productivity-centric context afforded by M365. Given the similarity between our intent taxonomy and that of \citet{chatterji2025chatgpt}, we are also able to compare user intent in M365 Copilot to that in ChatGPT.

The bulk of this paper characterizes the degree and nature of use of M365 Copilot across these intent and work activity classifications to support  two theses: first, that enterprise AI is becoming an everyday assistant for knowledge work; and second, that a possible next frontier is deeper adoption across underused work. Key findings:

\begin{itemize}
    \item \textit{M365 Copilot sees broad use across occupations and has areas for growth.} While some user intents such as \textit{Content Refinement} are frequent regardless of occupation, there is a long tail of usage types. Within that long tail, industry and occupational comparisons reveal work activities where the share of M365 Copilot is underrepresented compared to expectation based on the labor market generally. Some examples: the share of \textit{Evaluate the quality or accuracy of data} is lower than expected in the Banking industry; almost half of work activities done by employees in consulting are absent in M365 Copilot usage; the share of work activities involving \textit{Documenting/Recording Information} is lower than expected given the size of workforce that do those activities. Underrepresented usage indicates potential growth areas.
    \item \textit{M365 Copilot has unique usage compared to ChatGPT.} Writing dominates both platforms, but M365 Copilot sees more usage for decision making and problem solving, analyzing and processing information, getting technical help, and obtaining information, especially company-specific information. ChatGPT skews toward ideation in the form of creative thinking and interpreting information.
    \item \textit{Users may be shifting away from ``chat as search''.} We see a 5\% drop in the \textit{Information Inquiry} share of all user intent in our intent time series data. As shares of activity, this may be due to an increase in other types of usage rather than a decrease in the amount of information seeking done with M365 Copilot. Many occupations do more \textit{Content Refinement} than \textit{Information Inquiry} with M365 Copilot. 
    \item \textit{Use of M365 Copilot generally is assistive toward user goals.} Work activities that skew most toward the M365 Copilot side of user interactions center on advising and evaluating products, services, or technologies, assisting with paperwork and other services, and providing information for various purposes. 
    \item \textit{M365 Copilot targets information work.} This skew is expected. However, tracking the relevance of M365 Copilot to the full range of work can quantify the disparity and help identify emerging use cases. Our analyses of intent by occupational groups and job family ``AI applicability profiles'' quantify the gap in the relevance of AI to some classes of work over others, and provide views into potential growth of AI that may cross industrial boundaries. 
\end{itemize}

\section{Data \& Method} \label{sec:data}
\subsection{What is M365 Copilot}
Microsoft 365 Copilot is an AI-powered assistant embedded in Microsoft 365 applications that combines large language models with organizational data from Microsoft Graph, such as files, emails, and meeting transcripts, and web-grounded information for productivity tasks. Since its general availability for enterprises on November 1, 2023, Copilot has been available worldwide in Microsoft’s public clouds and is now used by nearly 70\% of the Fortune 500\footnote{Reported in Microsoft earnings on October 30, 2024.} and globally by more than a million companies. Based on direct classification of usage, we estimate that between 82\% - 89\% of M365 Copilot Chat usage is for work purposes, with about 10\% for personal purposes, a stark difference from the 27\% of ChatGPT usage for work \citep{chatterji2025chatgpt}. The user experience for M365 Copilot includes both stand-alone chat, app-embedded chat (in the side pane of Word, for instance), and on-canvas interaction (summarize an email thread in Outlook). All data used here were drawn only from M365 Copilot Chat, and thus does not include on-canvas in-application usage.
Because M365 Copilot is embedded in the applications where enterprise knowledge work already happens, usage logs offer a direct window into how AI is being woven into everyday work, rather than how people experiment with a stand-alone chatbot.

\subsection{Data}
We used five datasets, each derived from a sample of conversations between a user and M365 Copilot, all of which were performed securely using an ``eyes-off'' environment, meaning that no raw data was ever accessible directly for human review. Only anonymized and aggregated quantities, such as the percentages of each intent class, were egressed from the eyes-off data environment and only after thorough review and approval by Microsoft’s privacy and compliance processes. All aggregations were based on at least 5 users from 5 enterprise customer tenants, though the vast majority were derived from much larger samples. Finally, data subject to EU Data Boundary isolation was excluded from this analysis, as were data from educational institutions. We also aggregate our user intent datasets by Occupational Group, which consist of groups of users with job titles clustered using an LLM-based classifier. 
(see Appendix Figure \ref{fig:intentoccgroupall} for the full list of Occupational Groups). Occupation Group aggregations followed the same privacy and compliance approved processes as other data aggregations.

\textit{Intent dataset (primary)}: Our primary intent analyses were based on a sample of 175,274 M365 Copilot user prompts from the week of February 16 through February 22, 2026. The intent dataset was sampled randomly from the entire M365 Copilot Chat prompt corpus, aside from data subject to EU Data boundary isolation and educational institutions, including both free and paid users.

\textit{Intent dataset (time series)}: Our time trend dataset for user intent was compiled from a sample of approximately 310,000 conversations per week, sampled over a 114 day period from June 1, 2025 to September 22, 2025, yielding a total sample of 5,342,156 prompts. 

\textit{Work activity dataset (primary)}: Primary analyses of O*NET work activities were conducted on a dataset of 105,000 randomly sampled M365 Copilot conversations from global enterprise users, excluding data subject to EU Data Boundary restrictions and educational institutions. Sampling occurred at a rate of 15,000 conversations per day between February 14 and February 21, 2026. Each work activity analysis dataset included the entire Copilot conversation, encompassing both user prompts and Copilot responses.

\textit{Work Activity dataset (industry)}: To compare M365 usage across industries we sampled conversations from a selected set of 5-11 companies demonstrating significant M365 Copilot adoption with sustained user activity in each of three industries: \textit{Banking}, \textit{Discrete Manufacturing} (referred to throughout as \textit{Manufacturing}), and \textit{IT Services \& Business Advisory} (referred to throughout as \textit{Consulting}). Approximately 2,500 conversations per day were sampled randomly from each industry over the week February 14 – February 21, 2025, for a total of approximately 52,000 split evenly into 17,500 per industry. Sampling matched amounts across Banking, Manufacturing, and Consulting lets us compare how the same AI assistant is taken up in industries with different work mixes, which is central to identifying where enterprise AI is already established and where adoption still has room to grow.

\subsection{Intent Classification}
\subsubsection{Taxonomy Creation}
The user intent taxonomy consisted of two hierarchical levels, a top-level intent and secondary intents within each of the top-level intent classes. The intent classes were created by merging two sets of candidate intents, one learned by running the TnT-LLM taxonomy generation process \citep{wan2024tnt-llm} on the WildChat corpus of ChatGPT conversations~\citep{zhaowildchat}, and the second drawn from categories curated from product managers. The learned taxonomy from WildChat data provides bottom-up taxonomy creation, while the curated intents ensures relevance to M365 Copilot.
The final set of intents was used in a system prompt for classification. The system prompt was created and refined through an iterative process by continuing to provide correct and incorrect classification examples from WildChat data while the LLM classifier continued to classify the validation set more accurately without regressing on the test set.

\subsubsection{Classification}
The intent classification was based on one user prompt per M365 Copilot conversation. The specific user prompt chosen from a given conversation was selected at random but the LLM-based intent classification of that prompt included three other user prompts from the same conversation as context. This strategy supports random selection of the user prompt for classification rather than always using the initial user prompt, for example, while ensuring sufficient context should the selected user prompt be ambiguous in isolation (e.g., “give me 10 more”). The chosen user prompt was classified using the system prompt in two passes, first for top-level intents (e.g., \textit{Information Inquiry}, \textit{Content Generation}, \textit{Ideation and Planning}, \textit{Programming Assistance}, etc.)\ and then for secondary intents. GPT 4.1-mini was used for all intent classification.
    
\subsection{Work Activity Classification}
Work activity classification used the procedure developed by \citet{tomlinsonworking}, in which work activities detected in Copilot conversations are based on the O*NET classification maintained by the Department of Labor \citep{onet29}. O*NET is a hierarchical classification system that decomposes work done by occupations into tasks (most granular), detailed work activities, intermediate work activities, and generalized work activities (least granular). We classified M365 Copilot conversations based on the intermediate work activities (IWAs) they contain.

The IWA classification itself was performed in near identical fashion to \citet{tomlinsonworking}. A prompt (GPT-4.1-mini) and embedding-based (text-embedding-3-small) classification pipeline was used to identify all IWAs present in each M365 Copilot conversation, separately on the user (referred to as the \textit{user goal}) and Copilot (referred to as the \textit{Copilot action}) side of the conversation. In one minor departure from \citet{tomlinsonworking}, for performance reasons, when selecting the matched IWA from the set of potential candidates, we used only the top 20 most likely candidate IWAs. In accordance with \citet{tomlinsonworking} we set a frequency threshold of 0.05\% for IWA inclusion in all analyses to minimize the possibility of including falsely detected IWAs, where IWA frequency was determined by a \emph{fractional count} of IWAs detected in a given conversation. That is, for a conversation with $k$ matching IWAs, each matching IWA receives a fractional count of $1/k$, so the sum of all IWA fractional counts equals the number of conversations.

The O*NET taxonomy contains 332 IWAs. Table \ref{tab:iwadata} shows the count of fractional user goal IWAs and Copilot action IWAs, as well as the number of unique IWAs for each O*NET dataset after removing IWAs below the 0.05\% activity share threshold. In terms of non-fractional raw counts of IWAs, most conversations had more than one IWA on both the user goal and Copilot actions sides of the conversation. For our primary O*NET dataset, we see 176,803 absolute instances of user goal IWAs and 342,471 Copilot action IWAs. Thus in terms of sheer number of work activities detected in the overall sample, the number of work activities being performed by Copilot (3.26 per conversation) approaches double that of the work activities in user goals (1.68 per conversation).

\begin{table}[htb]
    \centering
    \caption{Summary of M365 Copilot datasets for work activity analyses}
    \label{tab:iwadata}
    \begin{tabular}{lccc}
        \textbf{Dataset} & \makecell{\textbf{User Goal, \# IWAs}\\(fractional count)} & \makecell{\textbf{Copilot Action, \# IWAs}\\(fractional count)} & \textbf{Unique IWAs} \\
        \hline
        Primary   & 70,068 & 86,361 & 169 \\
        Industry  & 52,286 & 58,271 & 178 \\
    \end{tabular}
\end{table}

We also classified the scope of the IWA being done and whether it was completed. Scope uses a 6-point Likert scale of none, minimal, limited, moderate, significant, and complete, with all analyses based on the percentage of matched conversations in which an IWA's scope was ‘significant’ or higher, the \emph{scope score}. Completion uses a 3-point Likert scale of not complete, partially complete, and complete, with all analyses based on the percentage of instances in which the per-IWA completion was equal to ‘complete,’ called the \emph{completion rate}. As our measures were identical to those in \citet{tomlinsonworking}, we refer readers to that work for validation, which was done with human raters or with explicit thumbs up/down feedback from users.

We used two measures related to work activity. First, we used activity share, which was the percent of total IWAs accounted for by each IWA using fractional share as defined above. If a dataset had a fractional count total of 100 IWAs and a given IWA had a fractional count total of 10, then its activity share would be 0.10. Second, we used the AI applicability score, as defined in \citet{tomlinsonworking}, which combines the activity share with the scope and completion rate as defined above. In our work we aggregated AI applicability scores from IWAs to job families and to GWAs as,
\begin{equation}\label{eq:ai-score}
    GWA_i^{\text{user}} = \sum_{j \in \text{IWAs}(i)}    \mathbf{1}[f_j^\text{user} \ge 0.0005]  c_j^\text{user}  s_j^\text{user} w_{ij},
\end{equation}
where IWAs$(i)$ is the set of IWAs mapping to GWA $i$,  $f_j^\text{user}$
is the user goal activity share of $j$,  $c_j^\text{user}$ is the task completion rate of conversations with IWA $j$ as a user goal, and $s_j^\text{user}$ is the fraction of conversations with user goal $j$ in which the scope classification is moderate or above. AI applicability for M365 Copilot actions is computed equivalently, and the overall AI applicability for an IWA, GWA, or job family is the average of the user goal and M365 Copilot action AI applicability scores.

\pagebreak \newpage
\section{Results} \label{sec:result}
\subsection{User Intent}
\subsubsection{Top-level Intent}
\begin{figure}[htb]
    \centering
    \includegraphics[width=0.95\linewidth]{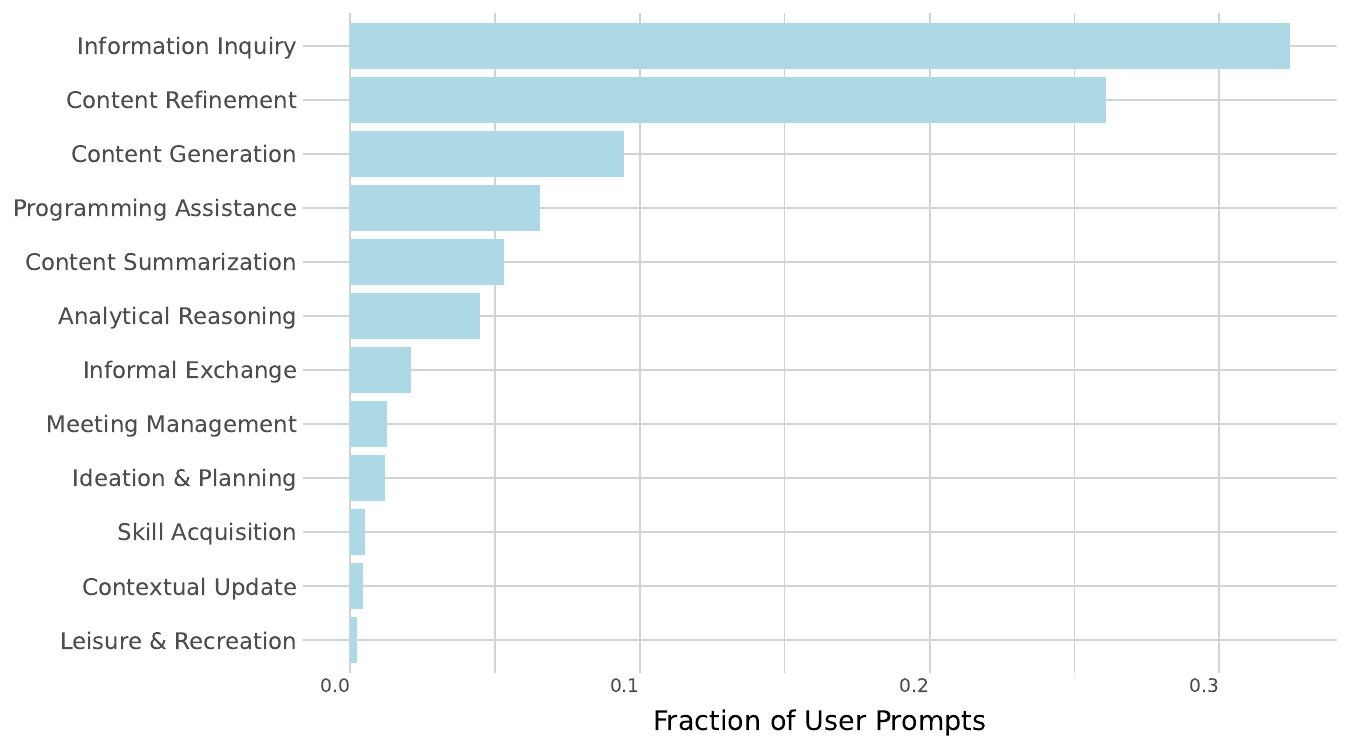}
    \caption{Fraction of user prompts for top-level intent}
    \label{fig:intentl1}
\end{figure}

Figure \ref{fig:intentl1} shows the distribution of user prompts in each top-level intent class. \textit{Information Inquiry} and \textit{Content Refinement} account for almost 60\% of user prompts. Other forms of content work including generation, as well as programming assistance and analytical reasoning account for about 20\% of user prompts. Content summarization, ideation and planning, and meeting-related activity account for about 7\% of usage, and the bulk of the remaining content (about 8\%) are ambiguous/other. Confirming the use of M365 Copilot for work purposes, only a little over 2\% of user prompts are for informal exchange or entertainment. Together, the concentration of usage in information inquiry and content work, and the near-absence of recreational use, are consistent with M365 Copilot functioning as an everyday assistant for enterprise knowledge work rather than as a general-purpose chatbot.

\subsubsection{Second-level Intent}
\begin{figure}[htb]
    \centering
    \includegraphics[width=0.95\linewidth]{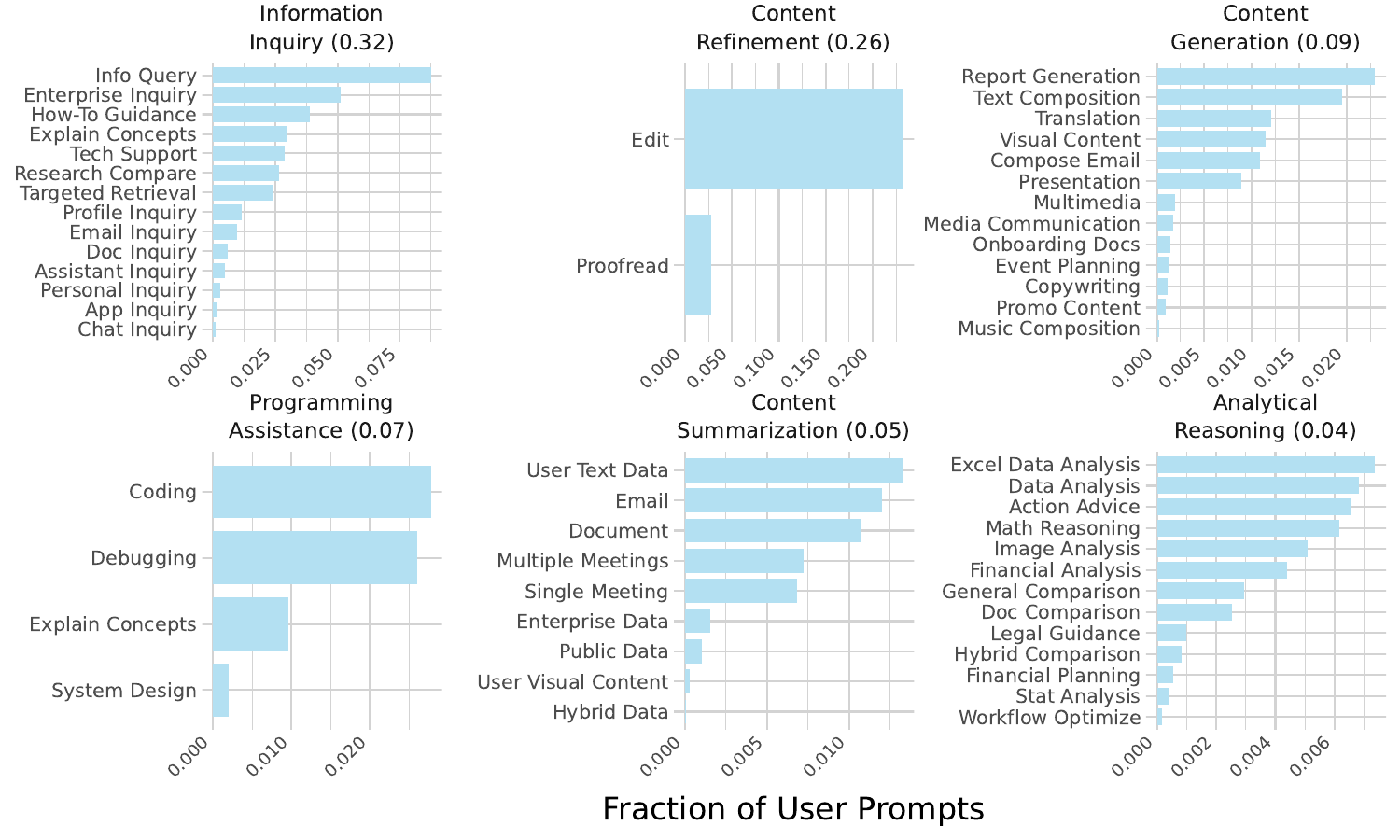}
    \caption{Second-level intents for each of the top 6 most frequent top-level intents. See Appendix Figure \ref{fig:intentl2all} for all second and first-level intent frequencies.}
    \label{fig:intentl2}
\end{figure}

For more detail, Figure \ref{fig:intentl2} shows the second-level intents for the six most frequent top-level intents. We observe that almost one quarter of all usage is for editing text for content refinement, followed by about 10\% of usage for asking general information questions. The distributions of the second-level intents underscore the broad, general purpose utility of M365 Copilot, spanning tasks from summarizing emails and files to specialized activities. The set of second-level intents that account for two percent or more of user prompts includes everything from asking for research and technical support to troubleshooting code to creating reports and PowerPoints.  Data analysis, including Excel data analysis, account for 1.5\% and image analysis account for about half of a percent of overall user prompts. In summary, while content rewriting or editing and general information question asking dominate with about one third of user prompts, the remaining two-thirds falls across a wide range of activities. 

\subsubsection{Intent Comparison to ChatGPT}
For comparison and generalization, we compare user intent in M365 Copilot to user intent in ChatGPT, as reported by \citet{chatterji2025chatgpt}. Because we do not use the same intent taxonomy, we manually mapped our second-level intent classes to the topics from OpenAI. For instance, \textit{Analyze Image} in our \textit{Analyze} top-level intent maps to \textit{Multimedia} in the OpenAI taxonomy, while \textit{Analyze Excel Data} maps to \textit{Technical Help}.

\begin{figure}[htb]
    \centering
    \includegraphics[width=0.95\linewidth]{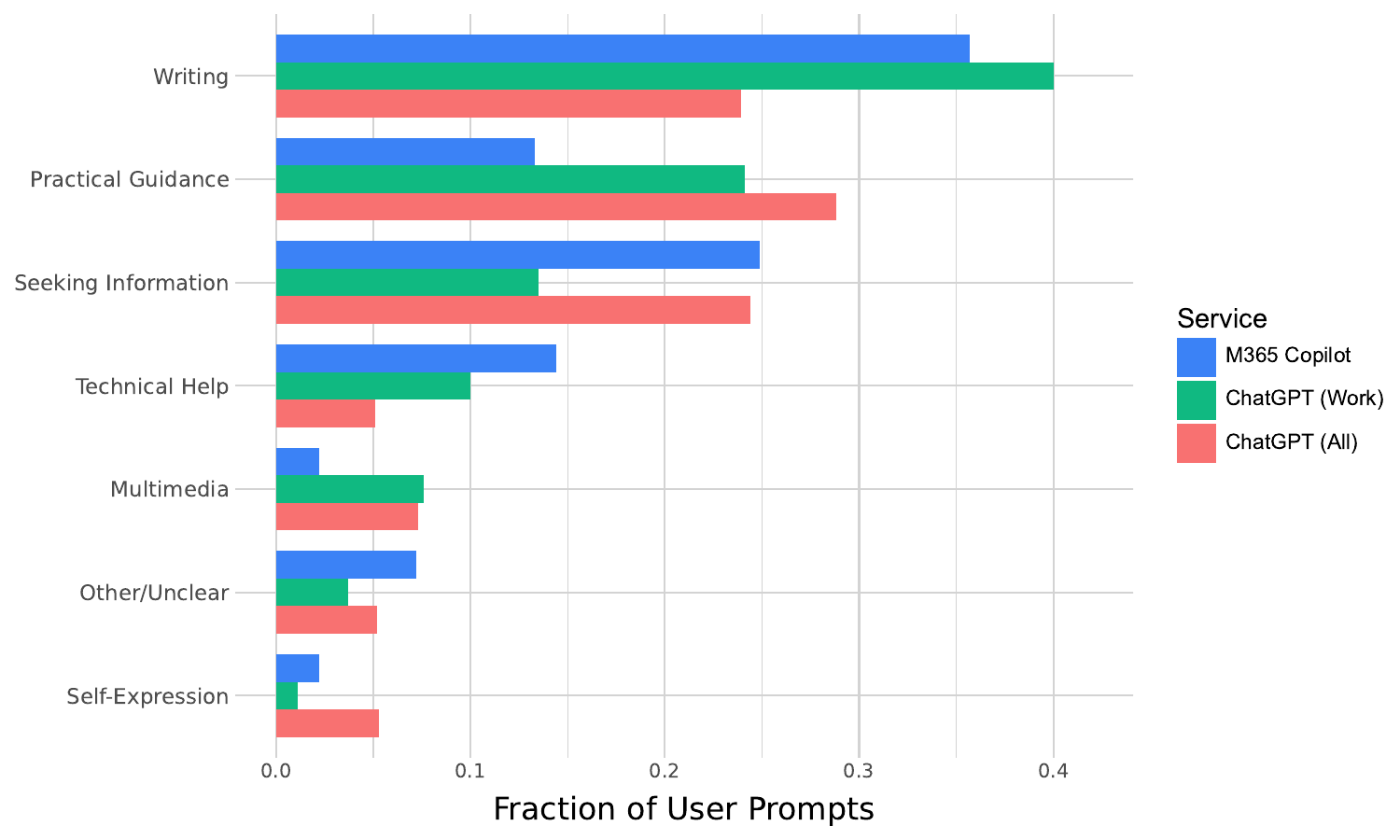}
    \caption{M365 Copilot and ChatGPT intents, mapped to the OpenAI topic taxonomy.} 
    \label{fig:intentl1chatgpt}
\end{figure}

Figure \ref{fig:intentl1chatgpt} shows the distribution of M365 intents mapped to the OpenAI topic taxonomy. When compared to the subset of ChatGPT for work-related purposes, both services exhibit a dominant share of \textit{Writing} as the most frequent user intent. Curiously, M365 Copilot and ChatGPT for work appear nearly reversed with respect to seeking information and practical guidance. This divergence may stem from M365 Copilot’s integration with organizational data, where company-specific information-seeking accounts for roughly one third of \textit{Seeking Information} intents, an aspect that is presumably minimally present in ChatGPT if at all. Furthermore, M365 Copilot has a higher proportion of \textit{Technical Help} (14\%) compared to ChatGPT for work (10\%). These differences with ChatGPT highlight usage of M365 Copilot as a more work-dedicated AI assistant. 

\subsubsection{Intent Over Time}
Given the rapid pace of adoption of AI (e.g., \citet{chatterji2025chatgpt,microsoftaiei2025,microsoftaidiffusionmethod}), how is usage changing as more people use, learn to use, and develop usage habits with AI? We look for changes in user intent over the 114 day period (June 1, 2025 - September 22, 2025) in our intent time series dataset. Importantly, these changes are relative to the overall sample of intent at each point in time and thus should be interpreted only as a distribution that evolves over time to reflect a changing balance of usage. \textit{Information Inquiry}, while still the most frequent intent, declined from 38\% of user intents to 33\%, a 13\% relative decline. In contrast, \textit{Content Refinement} rose from less than 22\% to 28.5\% for a relative increase of 32\%. Together, these two big shifts suggest users moving away from use of M365 Copilot for search and information lookup use and toward drafting and polishing of content. The generation of content itself has remained constant. 
This shift away from information lookup and toward content refinement is one of the clearest signals in our data that the role of M365 Copilot is deepening over time: users are moving beyond chat-as-search and toward using the assistant as a working collaborator in the production of content. Movement of this kind, away from the most basic interaction mode and into more substantive work, suggests that, as enterprise AI adoption progresses, the assistant becomes more    deeply embedded in day-to-day work.
Finally, we highlight a drop in the share of intent for \textit{Programming Assistance}, potentially attributable to increasing use of Github Copilot and other IDE-embedded AI tools, though attributing this decrease to such a shift requires confirmation. 

\begin{figure}[htb]
    \centering
      \includegraphics[width=0.95\linewidth]{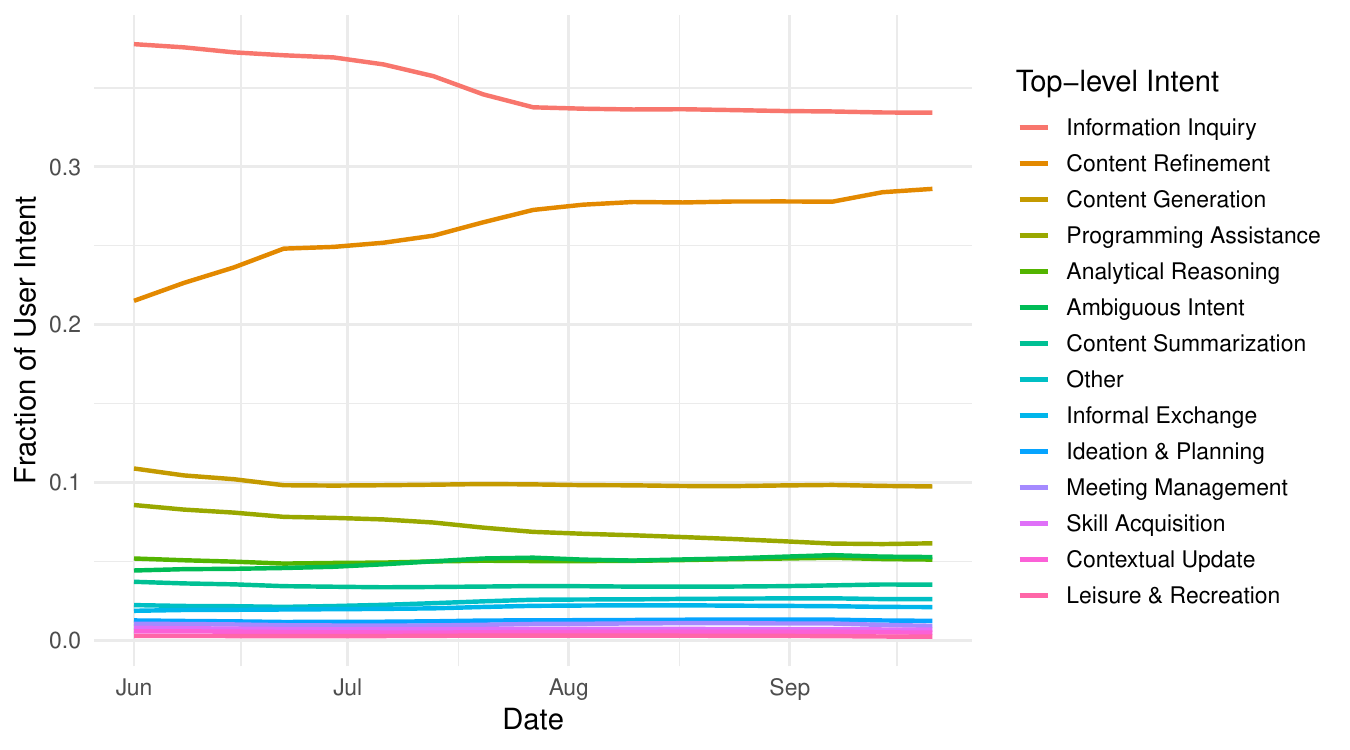}
    \caption{Fractional share of top-level intents over time. Data were sampled daily from June 1 - September 22, 2025. Values are averaged over rolling past 28 days.}
    \label{fig:intentl1time}
\end{figure}

\subsubsection{Intent by Industry and Occupation}
\begin{figure}[htb]
    \centering
    \includegraphics[width=0.95\linewidth]{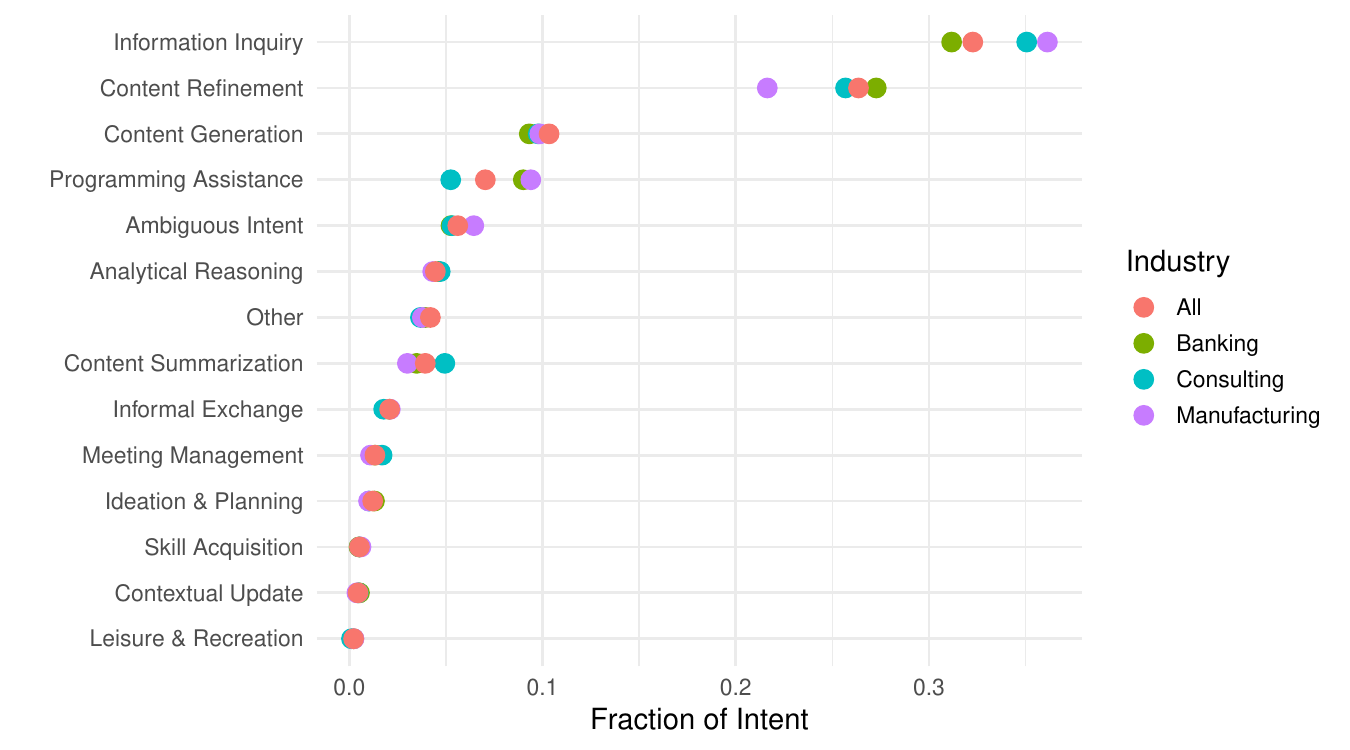}
    \caption{Top-level intents for three example industries, with the average across all 37 industries shown for reference. Data were sampled daily from Feb 16-22, 2026.}
    \label{fig:intentlindustry}
\end{figure}

We subset our intent data based on the originating company of each M365 Copilot prompt and aggregate to the same three example industries as in our O*NET industry dataset: \textit{Banking}, \textit{Manufacturing} and \textit{Consulting}. Figure \ref{fig:intentlindustry} shows that \textit{Information Inquiry} dominates, with more than 30\% of intent in each industry, though \textit{Banking} trails the all-industry mean, followed by \textit{Content Refinement}. 

\textit{Manufacturing} sits notably above the all industry mean for \textit{Information Inquiry} and below the all industry mean for \textit{Content Refinement}. \textit{Programming Assistance} is one intent where the industries are somewhat spaced out, especially relative to its share of all user intent. Here \textit{Manufacturing} (9.39\%) sees nearly double the share of \textit{Programming Assistance} as in \textit{Consulting} (5.25\%). We remind readers that these are relative shares. One industry could have both a larger absolute amount and smaller relative share of user intent in an intent class compared to another industry. 

\begin{figure}[htb]
    \centering
    \includegraphics[width=0.95\linewidth]{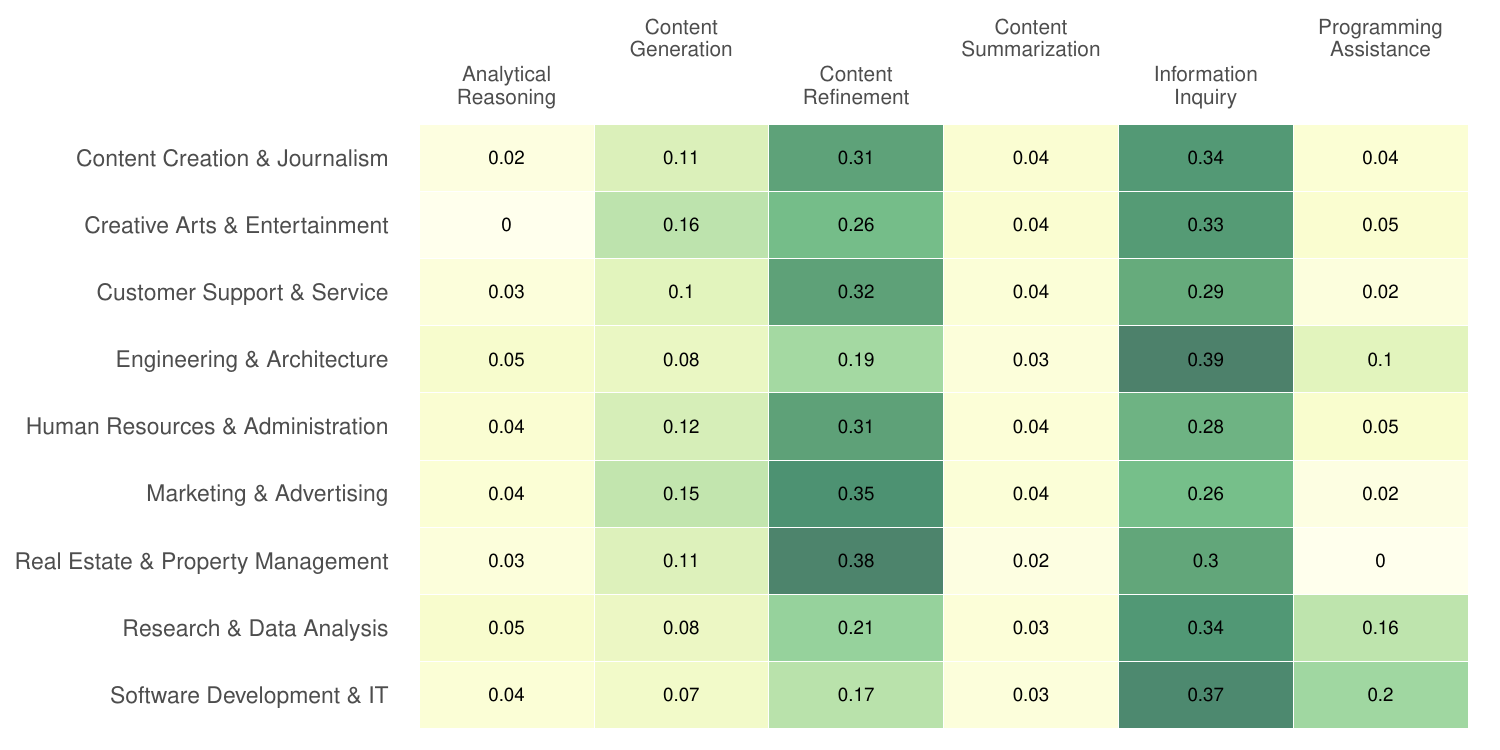}
    \caption{Share of top-level user intent by Occupational Group. For the full intersection of all Occupational Groups by all top-level user intents, see Appendix Figure \ref{fig:intentoccgroupall}.}
    \label{fig:intentjobfamily}
\end{figure}

We turn now to our examination of top-level user intent across Occupational Groups (Figure \ref{fig:intentjobfamily}). We see that some intents account for a larger share of usage (e.g., \textit{Content Refinement}) across all Occupational Groups, while others a lower share (\textit{Content Summarization}), highlighting functionality broadly applicable regardless of type of occupation. There are, however, many user intents with elevated relevance to specific Occupational Groups. \textit{Content Generation} stands out for \textit{Creative Arts \& Entertainment} (16\%), and \textit{Marketing \& Advertising} (15\%). \textit{Programming} has more than double the fraction of user intent  in \textit{Software Development and IT} and \textit{Research and Data Analysis} than the global average. A range of Occupational Groups are high in \textit{Content Refinement}, including \textit{Real Estate and Property Management}, \textit{Customer Support and Service}, \textit{Human Resources and Administration}, and \textit{Marketing and Advertising}. For all Occupation Groups crossed with all top-level intents, see Appendix Figure \ref{fig:intentoccgroupall}.

We generalize the analysis of relationships in the share distributions of the intents over Occupational Groups by clustering the intents. The resulting 4-cluster solution is shown in Table \ref{tab:intent_by_cluster}. Cluster 1 encompasses information, skill, and technical usage, Cluster 2 is about content work, Cluster 3 about meetings and staying current with communication and coordination, and Cluster 4 about skill acquisition and non-work.  

Together, these industry and occupational comparisons show the broad-but-uneven shape of enterprise AI adoption. Some intents, most notably \textit{Content Refinement}, are used heavily across nearly every occupation, suggesting that M365 Copilot is already serving as an everyday assistant cutting across occupational boundaries. At the same time, the modest inverse relationship between \textit{Information Inquiry} and \textit{Content Refinement}, and the sharp differences in programming-related use across industries, show that occupations and industries put the assistant to very different uses. Where an intent is common in some occupations or industries and rare in others, the gap itself marks a plausible direction for deeper adoption.
 
\begin{table}[ht]
\centering
\caption{Top-level intent clustered based on distribution of intent share over Occupational Groups.}
\footnotesize
\begin{tabular}{cccc}
\toprule
\textbf{Analytical} & \textbf{Content} & \textbf{Meeting} & \textbf{Social} \\

\midrule
\makecell[l]{Analytical Reasoning \\
Information Inquiry \\
Programming Assistance}
&
\makecell[l]{Content Generation \\
Content Refinement \\
Informal Exchange}
&
\makecell[l]{Content Summarization \\
Contextual Update \\
Meeting Management \\
Ideation \& Planning}
&
\makecell[l]{Skill Acquisition \\
Leisure \& Recreation}
\\
\bottomrule
\end{tabular}
\label{tab:intent_by_cluster}
\end{table}

\subsection{Work Activities}
\subsubsection{Generalized Work Activities}
\begin{figure}[htb]
    \centering
    \includegraphics[width=0.95\linewidth]{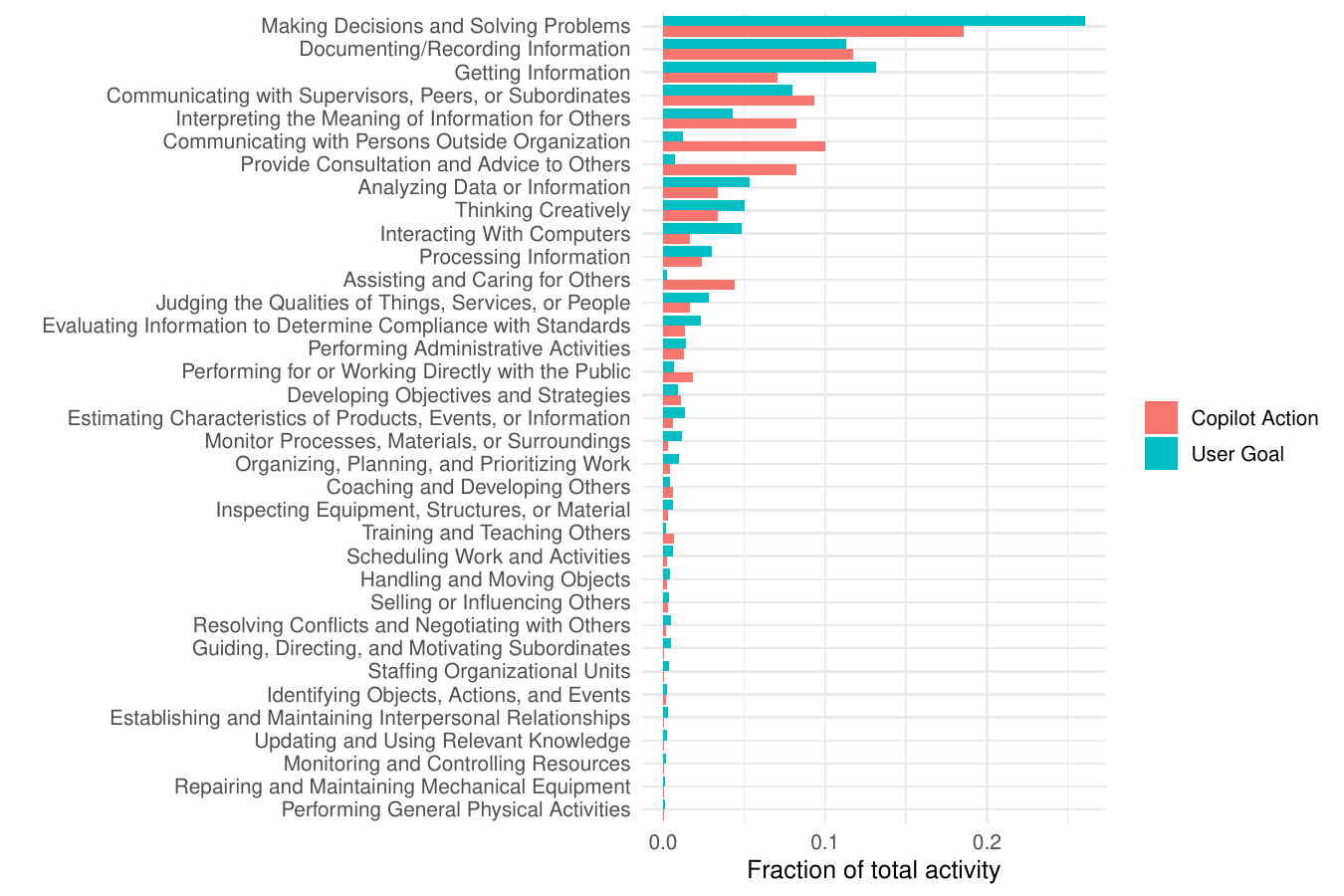}
    \caption{Fraction of each generalized work activity (GWA) for User Goal and Copilot Action}
    \label{fig:gwa}
\end{figure}

Our analysis of work activities starts with characterizing work activities performed with M365 Copilot through the aggregation of intermediate work activities (IWAs) into generalized work activities (GWAs) using the mapping provided by O*NET. This provides a comprehensive view of work done with and by M365 Copilot in a single set of the 35 (of 37 possible) GWAs seen in our data (Figure \ref{fig:gwa}). Recall that we separately classify the work activities on the user side of the conversation (user goal) and the M365 Copilot side (Copilot action). Top user goal GWAs center on information work, notably \textit{Making Decisions and Solving Problems,} \textit{Communicating with Supervisors, Peers, or Subordinates}, and \textit{Documenting/Recording Information}. Copilot actions show a similar skew toward information-centric tasks, with a stronger emphasis on interpreting information, providing consultation and supporting communication. 

\subsubsection{Intermediate Work Activities}
\begin{figure}[htb]
    \centering
    \subfigure[Top IWAs: User Goal]{
        \includegraphics[width=0.45\linewidth]{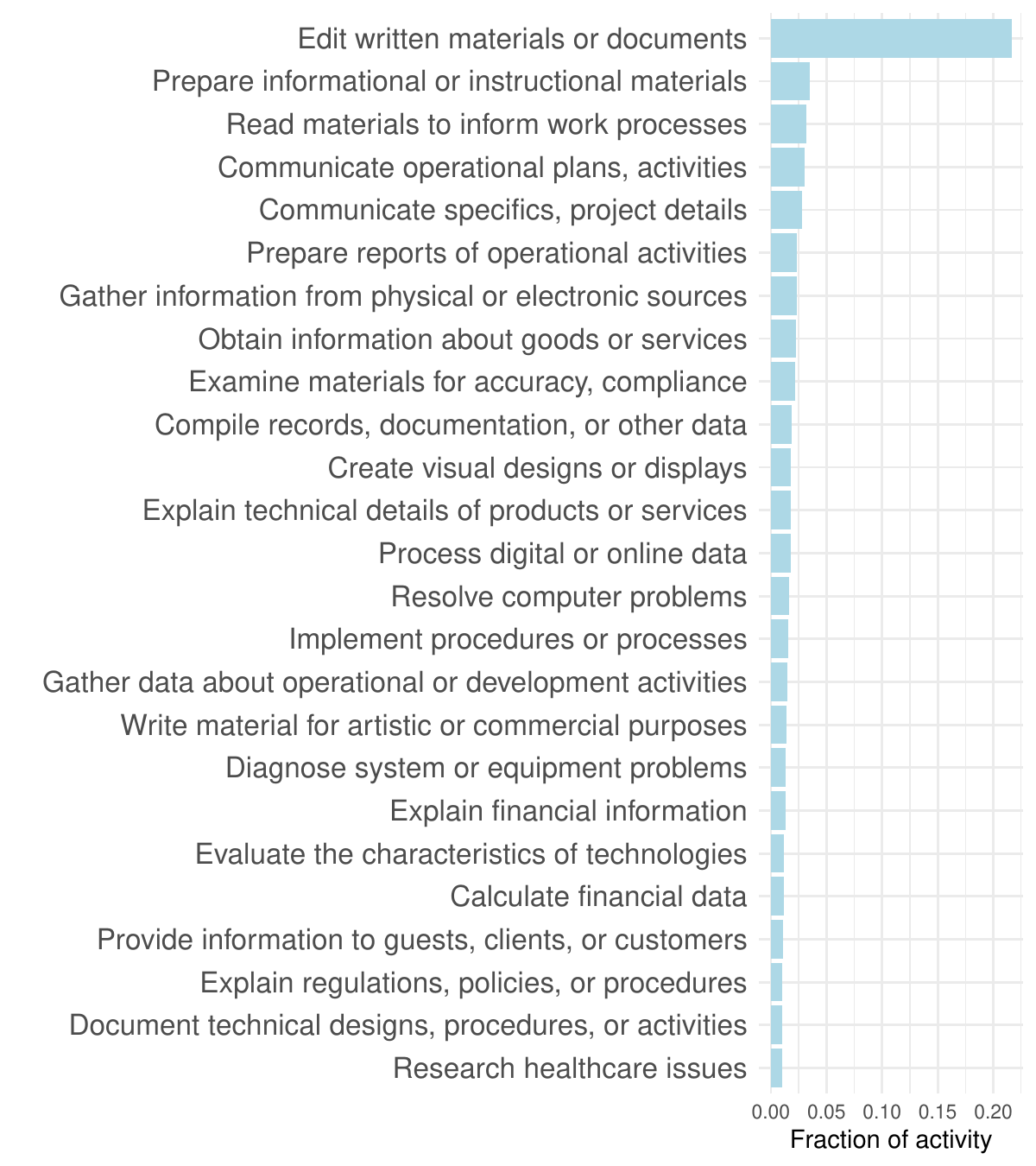}
    }
    \hfill
    \subfigure[Top IWAs: Copilot Action]{
        \includegraphics[width=0.45\linewidth]{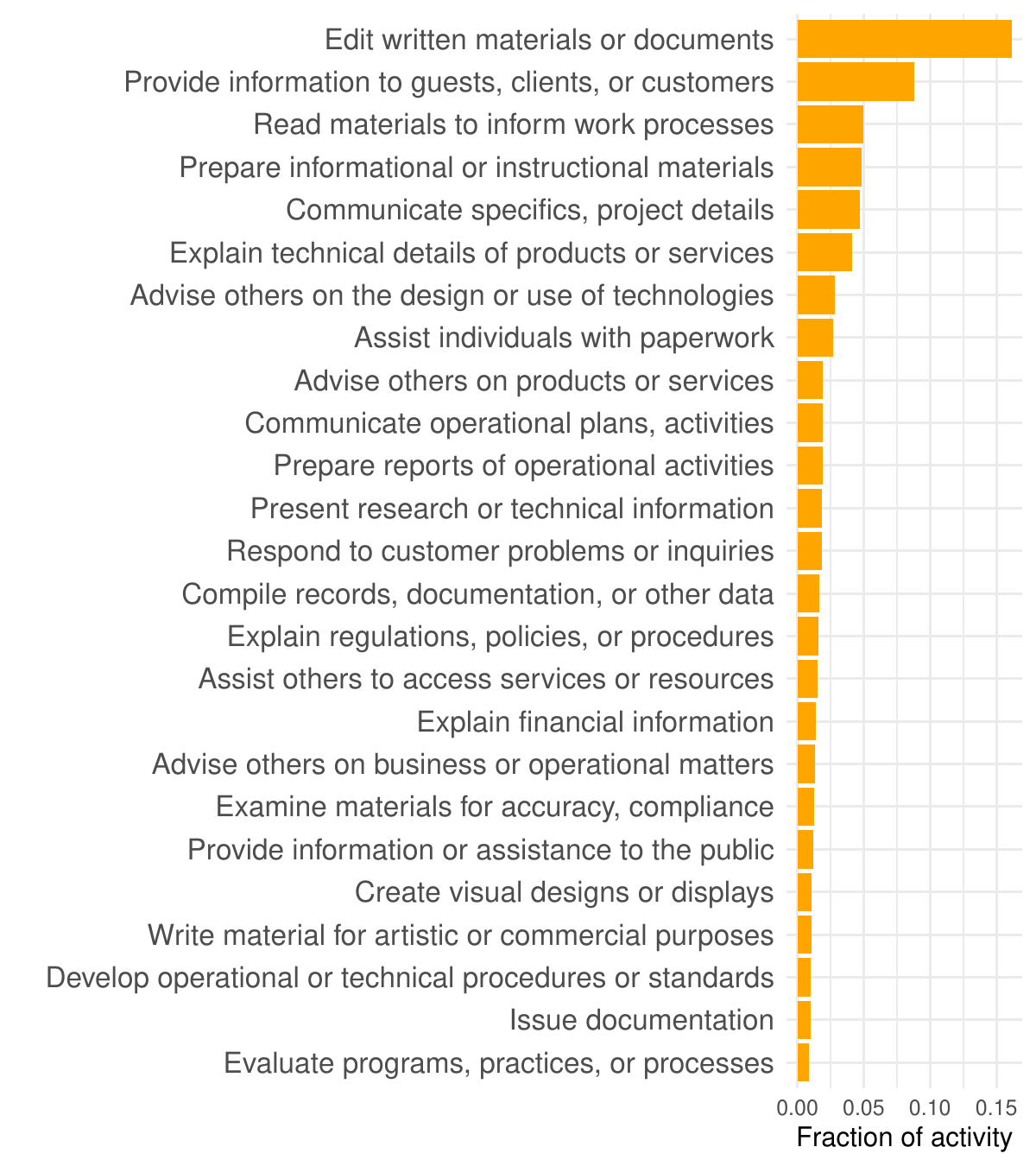}
    }
    \caption{Top 25 most frequent IWAs for user goal (a) and Copilot action (b). Note different x-axis scales.}
    \label{fig:iwa_top}
\end{figure}

We now examine intermediate work activities (IWAs) performed by users with Copilot assistance. Looking at the most frequent intermediate work activities (Figure \ref{fig:iwa_top}) done by users, writing and editing dominates, accounting for more than 20\% of activity. Other frequent IWAs reflect information work activities such as examining and preparing materials, communicating project details, gathering and analyzing data, diagnosing problems, and implementing procedures. Copilot actions appear largely complementary, with many top IWAs reflecting assistance, advice, and explanation given about technical details, regulations, and financial information.

The top 25 IWAs reflect ``head of curve'' information work tasks more common across occupations, but they don’t tell the whole story. In fact, these top 25 account for only two-thirds of all user goal IWA instances and three-quarters of Copilot action IWA instances. In other words, about 30\% of all work activities fall into the ``long tail'' of activities done with Copilot. Table \ref{tab:iwa_examples} shows example IWAs that are reasonably frequent but fall outside the top 25. The breadth of this long tail shows that users are already bringing a wide range of work activities to the assistant beyond the most frequent ones. It also points to where usage might grow. See Appendix Table \ref{tab:iwa_full_list} for a full list of IWA frequencies.

\subsubsection{User Goal and Copilot Action}

\begin{figure}[htb]
    \centering
    \includegraphics[width=0.95\linewidth]{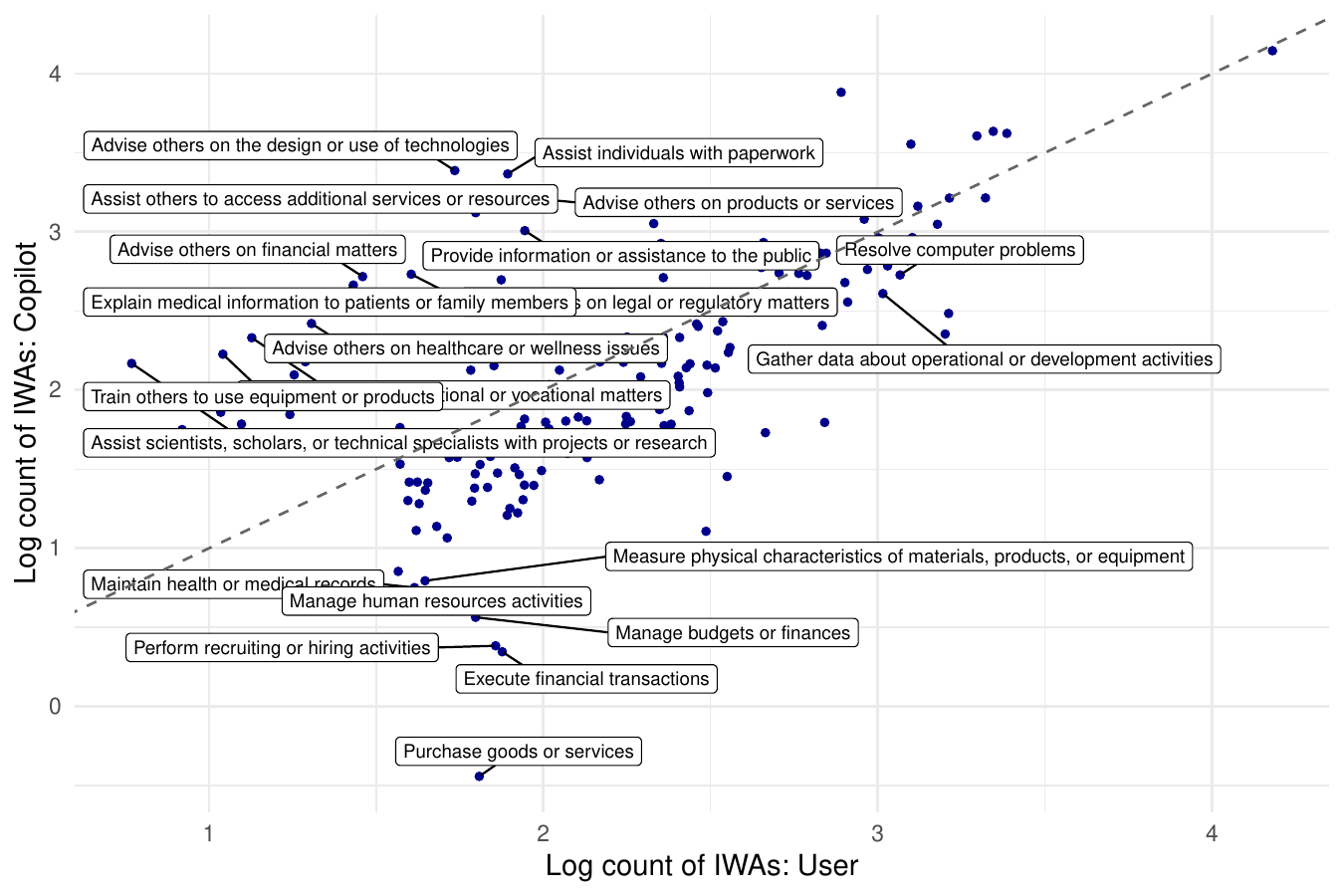}
    \caption{Frequency of all IWAs, split into user goal and Copilot action. Labeled IWAs indicate 10x more likely to be a Copilot action or 2x more likely to be a user goal.}
    \label{fig:iwa_scatter_user_copilot}
\end{figure}

In addition to the overall distribution of activities, we can also examine how the user side and the Copilot side of each conversation differ in their activity mixes. Returning to the question of the ways in which the goals of users deviate from the action taken by Copilot, Figure \ref{fig:iwa_scatter_user_copilot} highlights the IWAs that skew toward one or the other. User goals tend to emphasize data and information gathering, financial work, and administrative tasks, whereas Copilot actions predominantly focus on advising, assisting, and providing information.  The shape of this asymmetry — users bringing tasks and information needs, and Copilot responding with advice, assistance, and explanation — reflects Copilot's role as an assistant for knowledge work.

\begin{table}[htb]
  \centering
  \caption{Example IWAs outside the top 25 most frequent IWAs.}
  \label{tab:iwa_examples}
  \resizebox{\textwidth}{!}{
  \begin{tabular}{@{}cc|cc@{}} 
    \toprule
    \multicolumn{2}{c}{\textbf{User Goal}} & \multicolumn{2}{c}{\textbf{Copilot Action}} \\  
    \cmidrule(lr){1-2} \cmidrule(lr){3-4}
    \textbf{IWA} & \textbf{Fraction} & \textbf{IWA} & \textbf{Fraction} \\                        
    \midrule
    Assess regulations or policies & 0.009 & Evaluate programs, practices, or processes & 0.008 \\  
    Respond to customer problems or inquires & 0.007 & Develop marketing or promotional materials & 0.006 \\  
    Develop technical procedures or standards & 0.006 & Diagnose system or equipment problems & 0.007 \\  
    Analyze data to improve operations & 0.005 & Evaluate designs, specifications & 0.006 \\  
    Prepare financial documents & 0.005 &  Analyze business or financial data & 0.005\\  
    \bottomrule
  \end{tabular}
}
\end{table}

\subsubsection{Comparison to ChatGPT}
For broader context and to provide a sense of the uniqueness in usage of M365 Copilot as an exclusively productivity focused AI platform, we compare usage to that of ChatGPT \citep{chatterji2025chatgpt}. We draw on published ChatGPT usage reports, which enable near apples-to-apples comparisons of GWAs through common alignment with the O*NET taxonomy, though we note that the classification procedures and prompt structures used to classify ChatGPT data differ somewhat from those used for Copilot classifications. 

\begin{figure}[htb]
    \centering
    \includegraphics[width=0.95\linewidth]{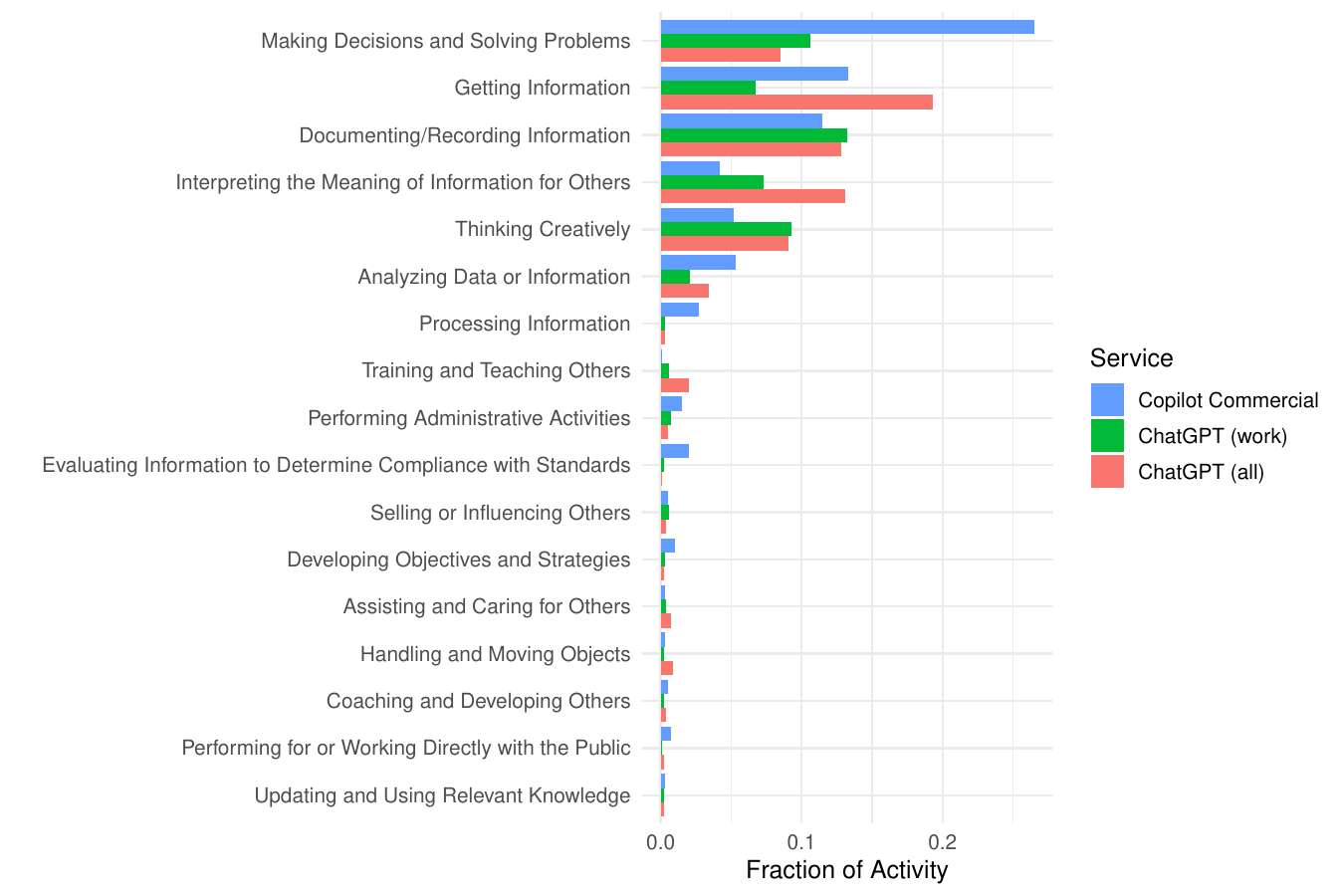}
    \caption{GWAs: M365 Copilot commerical and ChatGPT. Percents do not sum to 100\% due to limitation of publicly available ChatGPT data.}
    \label{fig:gwa_chatgpt_comm}
\end{figure}

In the higher-frequency GWAs shown in Figure \ref{fig:gwa_chatgpt_comm}, M365 Copilot is used disproportionately for \textit{Making decisions and Solving Problems}, as well as for getting, processing, and evaluating information, analyzing data, and developing strategies. By contrast, ChatGPT displays a relative emphasis on creative thinking and interpreting information. Both services are used heavily for documenting and recording information.

This divergence at the work-activity level mirrors the intent-level comparison: M365 Copilot is shaped by the analytical and decision-oriented work of enterprises, while ChatGPT skews more toward creative and interpretive uses.

\subsubsection{Labor Market, Industry, and Job Family Comparisons}
\begin{figure}[htb]
    \centering
    \includegraphics[width=0.95\linewidth]{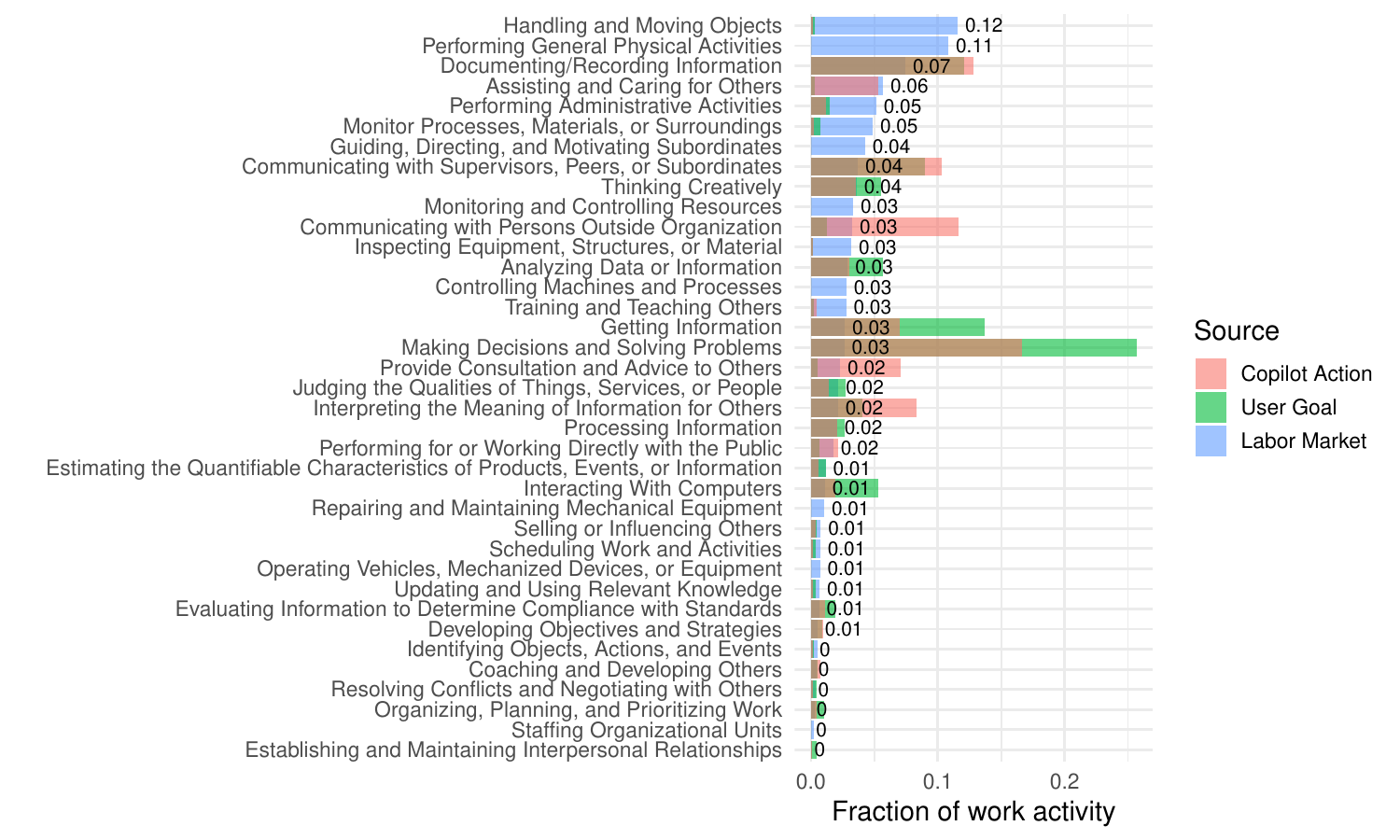}
    \caption{Fraction of generalized work activities (GWAs) in labor market. Fraction of user goal and M365 Copilot action skew toward information work.}
    \label{fig:gwa_user_copilot_market}
\end{figure}

We turn now to the coverage of M365 Copilot usage compared to work done in the labor market as a whole. Figure \ref{fig:gwa_user_copilot_market} shows the GWAs ordered by fraction of work in each GWA by the labor market overall. We can see that the top GWAs, \textit{Handling and Moving Objects} and \textit{Performing General Physical Activities}, which account for 23\% of all labor, see effectively no coverage by M365 Copilot. Conversely, GWAs such as \textit{Getting Information} and \textit{Making Decisions and Solving Problems}, which account for more than a quarter of the user goal GWAs account for only 6\% of labor done in the workforce at large. 

These differences naturally reflect the relevance and context of M365 Copilot as an information work technology. Given this expected divergence, for more detail, we plot the subset of IWAs seen in M365 Copilot usage, contrasting their activity share to their share in the labor market to highlight those that skew toward either the labor market or M365 Copilot (see Appendix Figure \ref{fig:iwa_user_copilot_market}a and Figure \ref{fig:iwa_user_copilot_market}b). 

\begin{figure}[htb]
    \centering
    \includegraphics[width=0.95\linewidth]{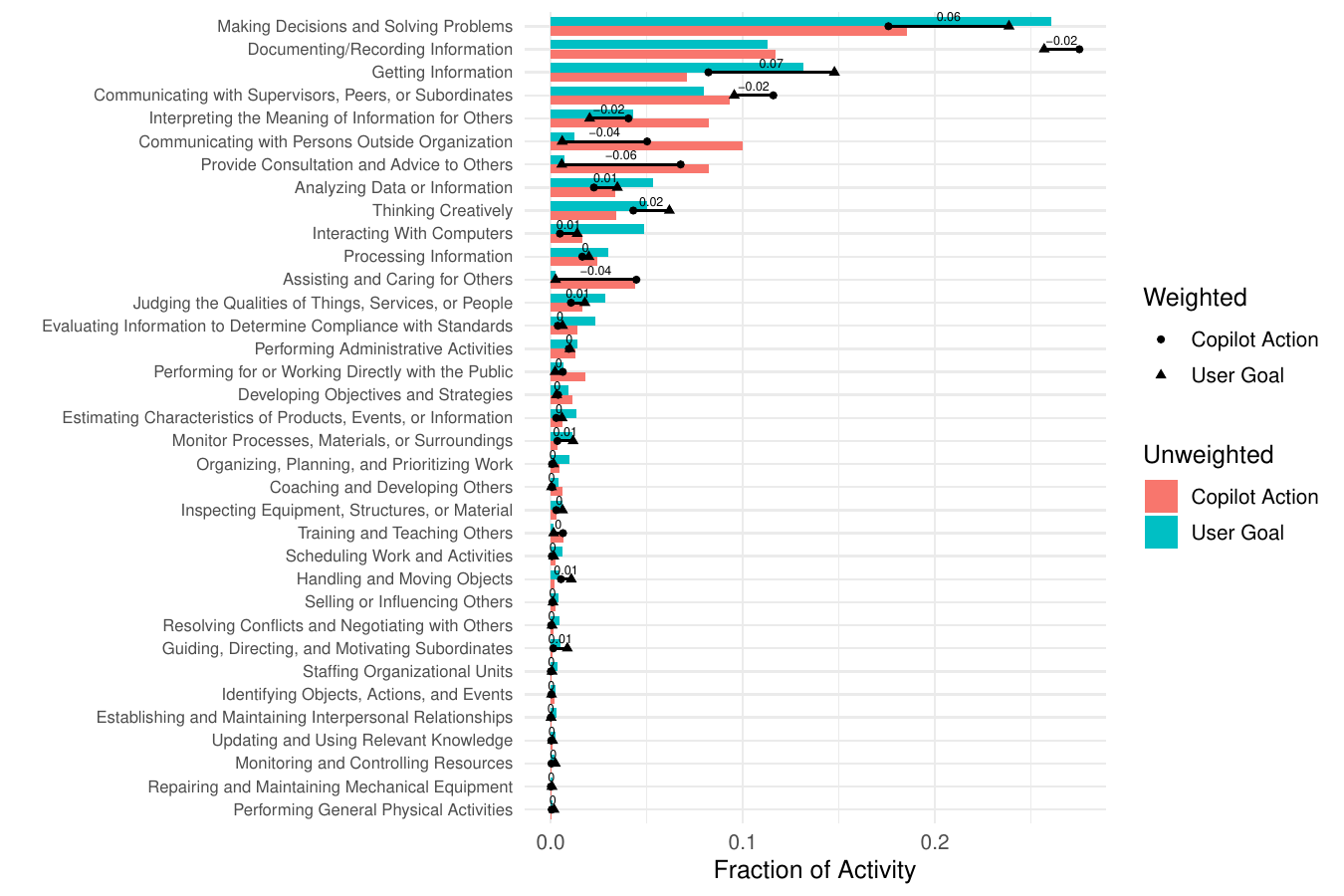}
    \caption{Weighted activity share for GWAs. Bars show activity share as in Figure \ref{fig:gwa}. Points show activity share weighted by workforce size.}
    \label{fig:gwa_activity_share_weighted}
\end{figure}

Even when restricting analysis to work activities involving M365 Copilot, we observe discrepancies between the activity shares in the data and those weighted by workforce size. In other words, comparing observed usage to expected usage based on the size of workforce that does each work activity reveals notable divergence (see Figure \ref{fig:gwa_activity_share_weighted}). Some work activities, such as \textit{Making Decisions and Solving Problems} see a slightly larger share than expected, though many are relatively underrepresented. For instance, while \textit{Documenting/Recording Information} is a common work activity conducted with M365 Copilot, we would expect even greater usage given the size of the workforce that perform those activities. \textit{Thinking Creatively} also sees a lower share than expected. Assuming M365 Copilot usage tends toward the amount of different types of work done, we would expect these underrepresented areas to grow going forward.

\begin{figure}[h]
    \centering
    \includegraphics[width=0.95\linewidth]{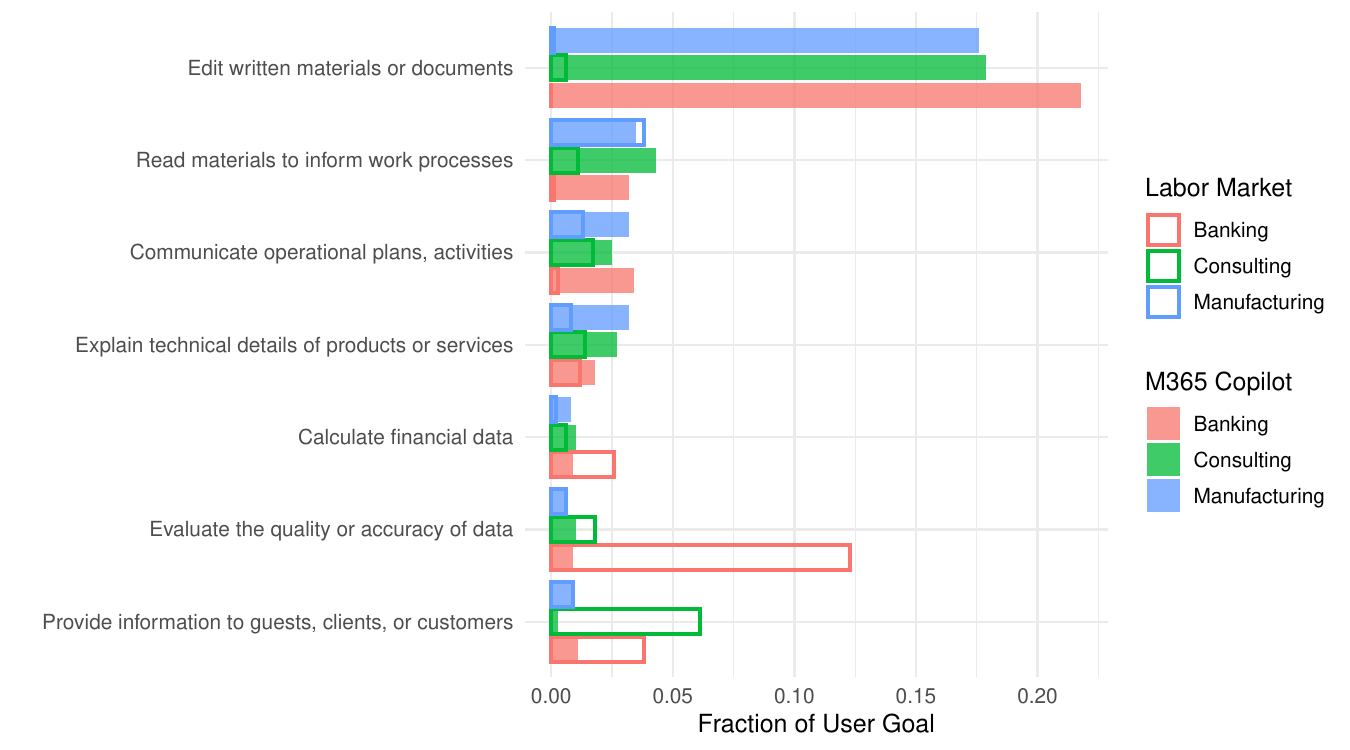}
    \caption{IWA comparison across three example industries. IWAs shown differ in M365 Copilot activity share by industry and/or from estimated share of work done in the industries based on breakdown of per-occupation workforce share within each industry taken from Occupational Employment and Wage Statistics. See Appendix Figure \ref{fig:iwa_industry_labor} for comparison across a larger set of IWAs.}
    \label{fig:iwa_industry_labor_ex}
\end{figure}

For a more detailed look at differences in M365 Copilot usage across the labor market, we can compare usage across industries. We illustrate by comparing estimates of IWAs done in each of the three selected industries (Banking, Consulting, and Manufacturing) to the share of corresponding IWAs seen in M365 Copilot. The fraction of work done in each industry was estimated using Occupational Employment and Wage Statistics data \citep{oews2024}. For each industry, the fraction of each IWA done was estimated as a sum of the IWAs performed by each occupation in the sector, weighted by occupation size. Per occupation IWA frequency was itself estimated as a weighted sum of task frequency as reported in O*NET. Noting that many IWAs in each industry are not present in the M365 Copilot data (Table \ref{tab:iwa_coverage}), Figure \ref{fig:iwa_industry_labor_ex} shows the fraction of user goal for each industry in a subset of IWAs that vary notably across the industries or in relation to estimated work done by employees in those industries.

\begin{table}[htb]
\centering
\scriptsize
\caption{Industry IWAs not present in M365 Copilot}
\begin{tabularx}{\linewidth}{l r r r l}
\toprule
\textbf{Sector} &
\textbf{Total IWAs} &
\makecell{\textbf{IWAs not in}\\\textbf{M365 Copilot}} &
\makecell{\textbf{Estimated}\textbf{ Work in}\\\textbf{non-present IWAs}} &
\makecell{\textbf{Example}\textbf{ non-present IWA}} \\
\midrule
Banking        & 238 & 140 & 36.6\% & Negotiate contracts or agreements \\
Consulting     & 315 & 217 & 46.9\% & Evaluate production inputs or outputs \\
Manufacturing  & 295 & 199 & 56.7\% & Assemble products or work aids \\
\bottomrule
\end{tabularx}
\label{tab:iwa_coverage}
\end{table}
Several IWAs stand out as divergent in M365 Copilot usage compared to the estimate of work done in the industry, including in particular \textit{Edit Written Materials or Documents}, which captures a far larger fraction of activity in M365 Copilot than in the industries themselves. This discrepancy is due, presumably, to the general usage of M365 Copilot for editing and writing in the course of many jobs, while each industry likely has few jobs, such as copy editing, where editing and writing constitute ``official'' work activities in O*NET.

Some lower frequency work activities in M365 Copilot may undershoot potential usage, particularly in Banking where activities like \textit{Evaluate the quality or accuracy of data} are far below their estimated overall share. These suggest areas where fine tuned AI agents may be useful for assisting with work. Evaluating data in the banking industry, for instance, is a work activity commonly done by claims adjusters and insurance underwriters who may have specific workflows, tools, or data analysis needs that require more customized AI solutions.

\begin{figure}[htb]
    \centering
     \includegraphics[width=0.95\linewidth]{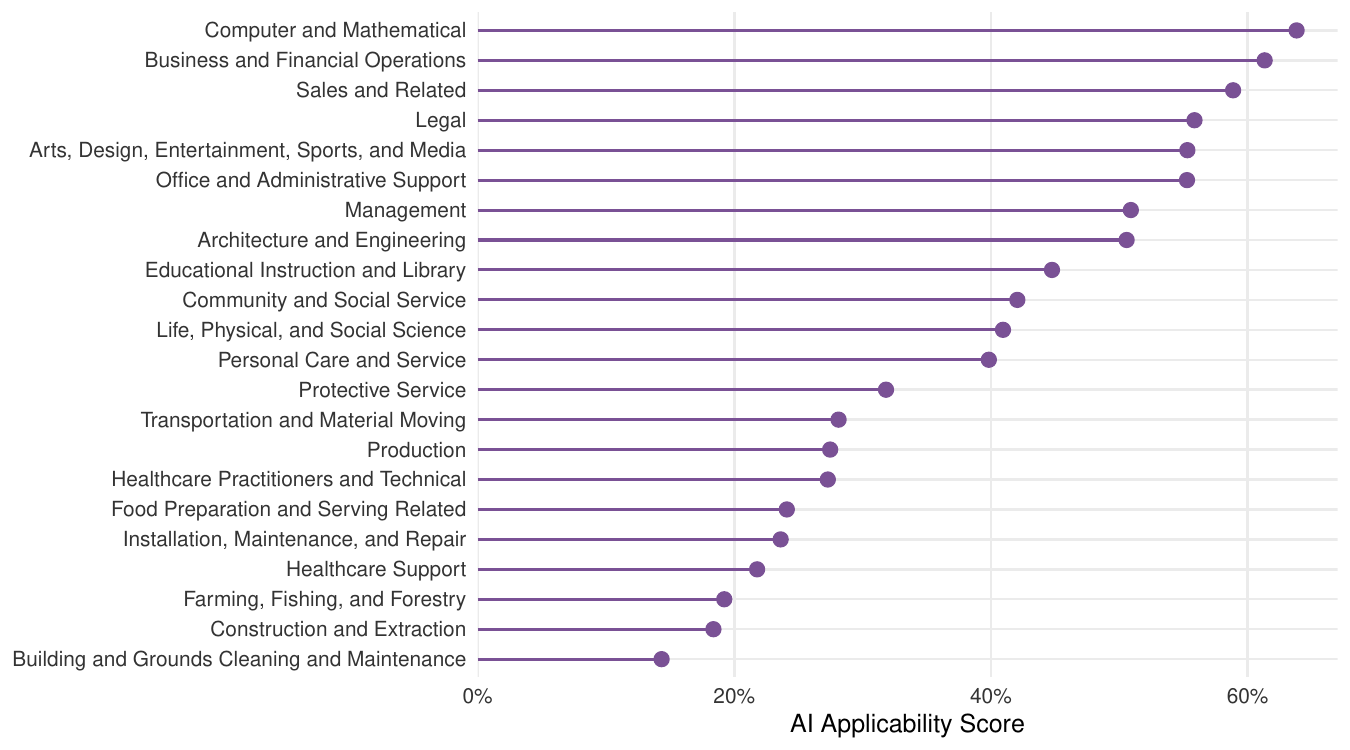}
    \caption{AI applicability score by O*NET job family.}
    \label{fig:job_family_ai_applicability}
\end{figure}

Next we apply the AI applicability metric proposed by \citet{tomlinsonworking}, aggregated at the job family level as defined by O*NET. The AI applicability score, computed on IWAs and then aggregated to occupations, indicates the extent to which work activities seen in M365 Copilot usage are relevant to an occupation. 
Figure \ref{fig:job_family_ai_applicability} highlights the relevance of M365 Copilot to information work occupations, led by the \textit{Computer and Mathematical} job family. In broad strokes, two groups of job families can be seen in Figure \ref{fig:job_family_ai_applicability}, one for information work occupations, where the AI applicability score is 40\% or higher, and one for physical world occupations where the AI applicability score is 25\% or lower.

To understand the makeup of work done in each job family we decompose the AI Applicability score for each job family into constituent GWAs. This allows us to understand not just the overall degree of AI applicability to each job family, but a ``profile'' of the ways in which AI is applicable based on usage. Although the \textit{Administrative Support} and \textit{Art, Design, and Media} job families, for instance, share the same AI applicability score of 0.55, the composition of this metric by GWAs differs substantially. \textit{Art, Design, and Media} has far more \textit{Thinking Creatively} while \textit{Administrative Support} sees much more \textit{Processing Information}, \textit{Administrative Tasks}, and \textit{Communicating with Externals}.

\begin{figure}[htb]
    \centering
    \subfigure[Computer and Mathematical]{
         \includegraphics[width=0.45\linewidth]{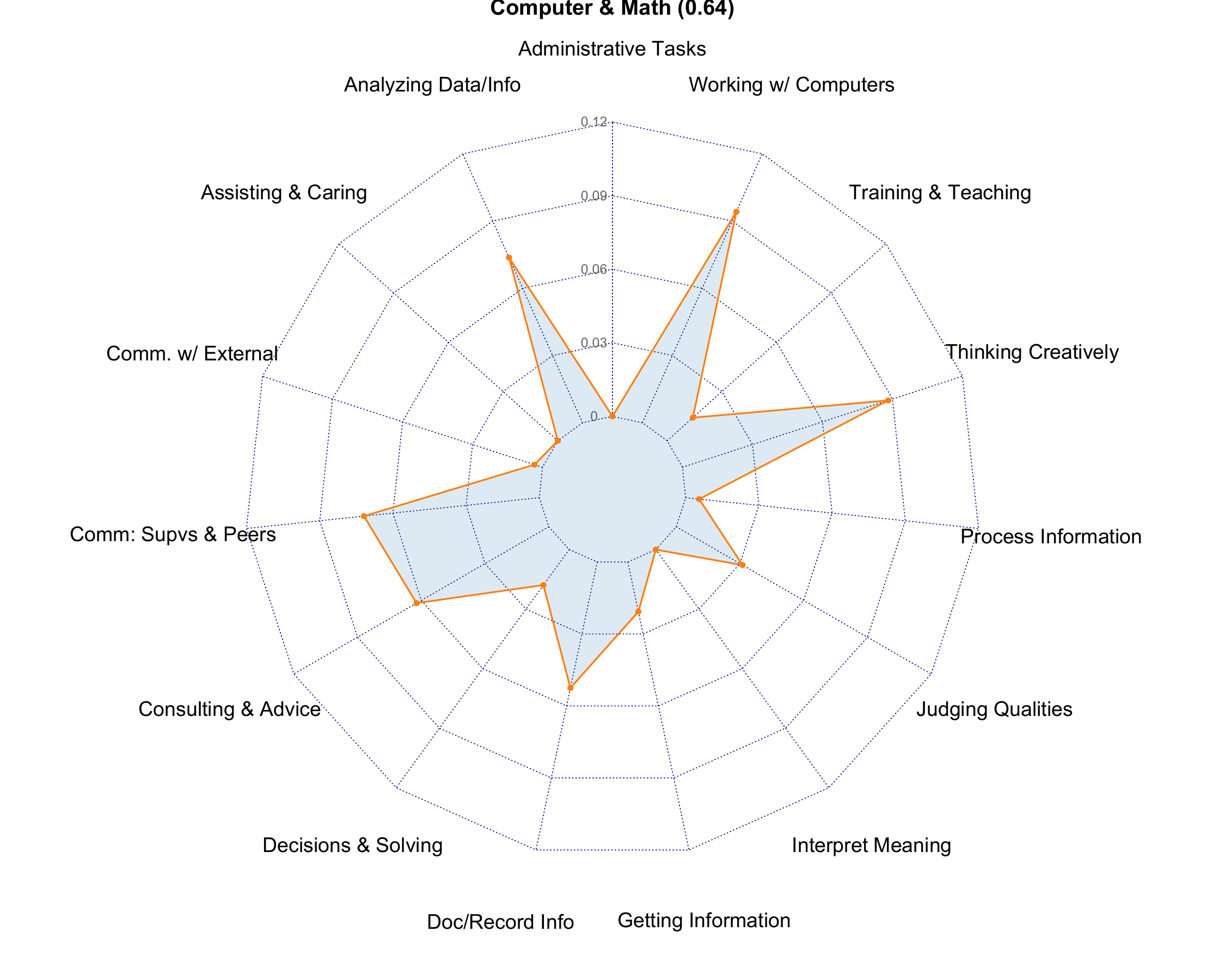}
    }
    \hfill
    \subfigure[Educational Instruction and Library]{
        \includegraphics[width=0.45\linewidth]{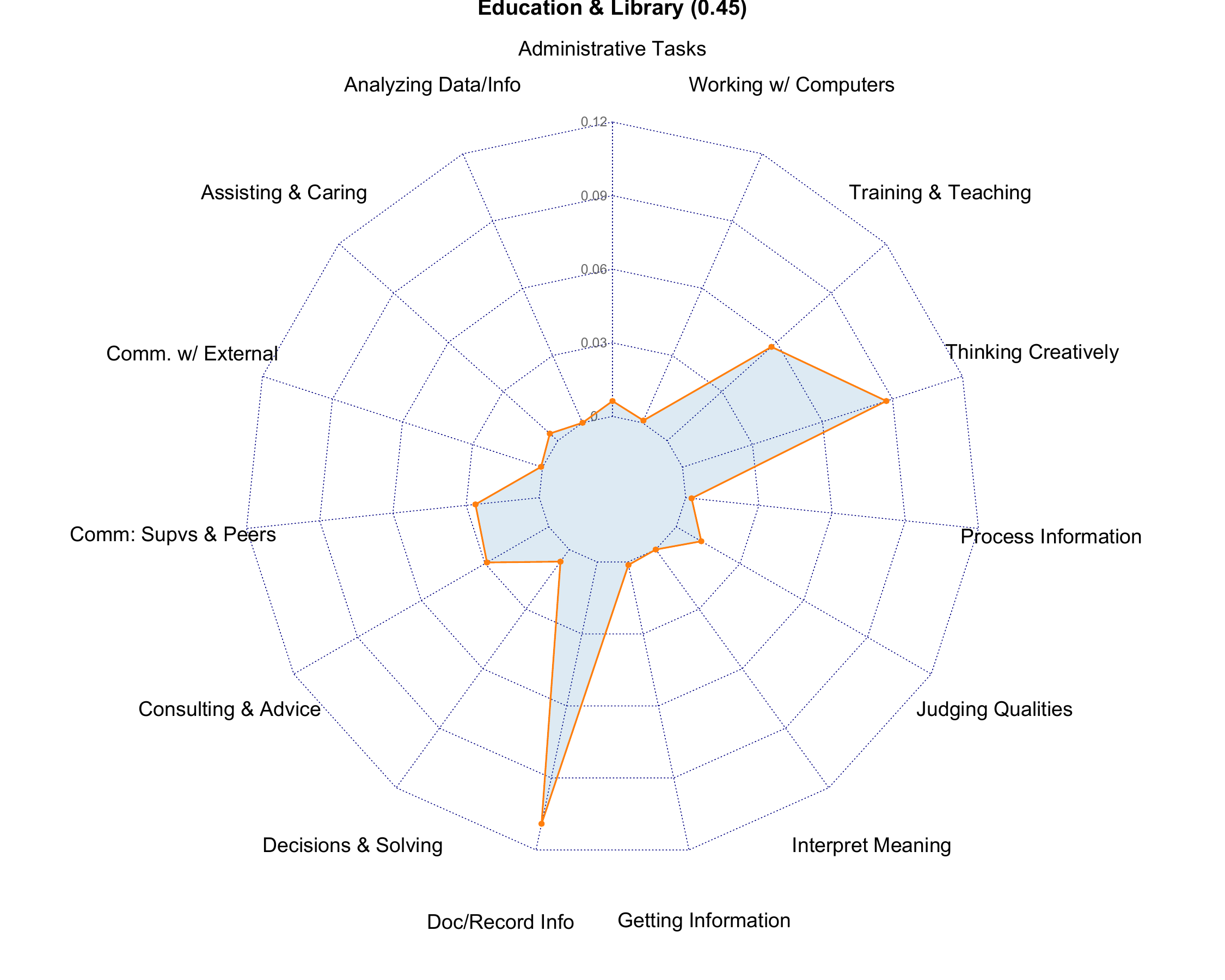}
    }
    \caption{AI applicability score and breakdown by GWA for the Computer and Mathematical (a) and Educational Instruction and Library (b) job families. See Figure \ref{fig:job_family_radar_all} for more job families.}
    \label{fig:job_family_examples}
\end{figure}

We illustrate these job family profiles in Figure \ref{fig:job_family_examples}. \textit{Computer \& Mathematical} has a somewhat higher overall AI applicability score (0.64) than \textit{Educational Instruction and Library} but the differences in how that applicability is distributed stand out. Both see considerable \textit{Thinking Creatively}, but \textit{Educational Instruction and Library} sees more applicability in \textit{Training \& Teaching}, \textit{Judging Qualities}, and \textit{Documenting/Recording Information}. \textit{Computer \& Mathematical} sees more AI applicability in \textit{Analyzing Data and Information}, \textit{Working with Computers}, among other areas largely absent for \textit{Educational Instruction \& Library}.

Job families with lower AI applicability scores can also be differentiated. \textit{Life \& Social Sciences} is spiky, with considerable applicability in \textit{Analyzing Data \& Information} and \textit{Documenting \& Recording Information}. \textit{Personal Services} is only moderately lower in overall AI applicability, but shows a more well-rounded usage pattern across work activities. See Appendix Figure \ref{fig:job_family_radar_all} for AI applicability profiles for more job families.

Taken together, these three views — labor market, industry, and job family — give the most concrete picture in our data of one of enterprise AI's next frontiers: extending into the work activities, industries, and occupations where there is the most room for adoption to grow.

\section{Discussion} \label{sec:discussion}

The findings above suggest two things about the state of enterprise AI today. First, M365 Copilot is functioning as an everyday assistant for knowledge work, used broadly across occupations and industries for activities ranging from writing and information lookup to analysis, decision support, and communication. Second, that adoption, while broad, is uneven. Usage patterns vary substantially across industries, occupations, and work activities, and there is substantial room for adoption to grow across the activities done in the labor market. For example, the distribution of top-level user intent shows both intent-driven effects in which some intents (\textit{Content Refinement}) comprise a far larger share of user intent than other intents (\textit{Meeting Management}) across all occupational groups. Most occupational groups do, however, have one if not several intents with uniquely higher share. \textit{Marketing \& Advertising} is relatively higher in \textit{Ideation \& Planning}, \textit{Science \& Research} higher in \textit{Analytical Reasoning}, and \textit{Software Development \& IT} higher in \textit{Programming Assistance}.

We see AI applicability scores for some information work job families (e.g., \textit{Business and Financial Operations}) more than double job families comprised of occupations dominated by physical world work (e.g., \textit{Construction and Extraction}). This skew away from physical labor and toward information work is not surprising, but tracking the uptake of AI across the labor market broadly is important for characterizing the evolving ``jagged edge'' of AI relevance. Our analysis of work activities across job families, shows that each job family is unique in the distribution of work done with M365 Copilot, even compared to job families with similar overall levels of AI applicability. \textit{Business \& Finance} looks different from \textit{Legal}, which looks different from \textit{Sales}. Comparisons across three industries corroborated this pattern: broad similarity in the work activities that comprise the most usage but with some uniqueness, especially in relative terms within less frequent work activities. 

What, then, does this uniqueness across industry and occupational comparisons suggest about directions and work contexts in which AI usage may grow? We identify two types of potential growth areas: 1) work done in the world and not present in M365 Copilot data, and 2) work done in the world but relatively underrepresented in M365 Copilot data. Our industry comparisons highlight the former case, as half or less of the work activities done in our example industries were present in the M365 Copilot data. While some of these, especially those oriented toward physical labor simply may not be a good fit for current generative AI, the industries selected largely are information work industries and thus comprised of work areas that are candidates for increased AI usage as AI capabilities improve and become embedded in and tailored to an increasing number of professions.

The industry comparisons also highlight areas of relative underrepresentation of AI usage. \textit{Banking} and \textit{Consulting} see about the same share of M365 Copilot usage for \textit{Evaluate the quality or accuracy of data}, but the banking industry as a whole does far more of that work activity, and thus we may see growth of that use case within that industry provided sufficient AI functionality. The analyses of job families show a similar pattern. Contrasting GWAs with and without workforce weighting highlights types of usage that may grow given the share of the labor market that does those activities. As examples, \textit{Thinking Creatively} and \textit{Communicating with Supervisors, Peers, or Subordinates}, while already common work activities done with M365 Copilot would see even higher usage share if that share were more aligned with the share of the workforce that does those activities.

Turning to the nature of usage, we see that high-frequency usage of M365 Copilot is consistently assistive via general-purpose functionality applied by people across a wide range of job types. The top user intent is still Information Inquiry and similarly Getting Information accounts for the second largest share of work activity on the user goal side. Writing is a dominant work activity, and both information seeking and writing are tasks done by people in many jobs rather than only by those whose primary work functions involve them. Returning to the industry comparisons, we see evidence of this: the share of people working in those industries with core work functions that involve writing is tiny compared to the share of writing done with M365 Copilot. 

Notably, despite the large share of M365 Copilot activity involving some form of information seeking, that was the area that saw the largest drop in usage over our six month time frame. AI adoption and the underlying technology and implementations are changing rapidly, but six months arguably is a short time frame to see the 5\% absolute drop in information seeking as a share of total usage. Such a shift may mark a turning point as people move away from using AI largely as a search engine replacement toward a wider variety of uses. We caution that we were only able to look at time trends for intent over a 114 day period so we raise this as suggestive but important, and recommend continued study to track this potential change. 

What emerges across these analyses is a picture of enterprise AI past the novelty stage and into something more substantive: M365 Copilot is being used as an everyday assistant across the occupations and activities of day-to-day knowledge work. At the same time, the pattern of where it is used, and where it is not, makes clear that today's usage is not the ceiling. Across industries, occupations, and the work the labor market actually does, sizable areas remain where there is room for adoption to grow. The next phase of enterprise AI, whatever else it brings in capability, will be defined in part by how broadly usage extends into those areas — and continued direct measurement of usage in real workflows is what will let us see whether and how that happens.

\clearpage
\setlength\bibsep{0.5\baselineskip}
\bibliographystyle{plainnat}
\bibliography{references}

@misc{onet29,
  author       = {{National Center for O*NET Development}},
  title        = {{O*NET Database Version 29.0}},
  year         = {2024},
  url          = {https://www.onetcenter.org/db_releases.html},
  note         = {Accessed: 2025-05-29}
}

@misc{oews2024,
  author       = {{U.S. Bureau of Labor Statistics}},
  title        = {{Occupational Employment and Wage Statistics (OEWS), May 2024}},
  year         = {2024},
  url          = {https://www.bls.gov/oes/tables.htm},
  note         = {Accessed: 2025-05-29}
}

@misc{gallup2025,
  author       = {Ryan Pendell},
  title        = {{AI Use at Work has Nearly Doubled in Two Years}},
  year         = {2025},
  url          = {https://www.gallup.com/workplace/691643/work-nearly-doubled-two-years.aspx},
  note         = {Accessed: 2025-10-19}
}

@article{bick2024rapid,
  title={The rapid adoption of generative {AI}},
  author={Bick, Alexander and Blandin, Adam and Deming, David J},
  journal={Management Science},
  year={2026},
  publisher={INFORMS},
doi={10.1287/mnsc.2025.02523}
}

@article{eloundou2024gpts,
  title={{GPTs are GPTs}: Labor market impact potential of {LLMs}},
  author={Eloundou, Tyna and Manning, Sam and Mishkin, Pamela and Rock, Daniel},
  journal={Science},
  volume={384},
  number={6702},
  pages={1306--1308},
  year={2024},
  publisher={American Association for the Advancement of Science}
}

@misc{handa2025economic,
  title={Which economic tasks are performed with {AI}? Evidence from millions of {Claude} conversations},
  author={Kunal Handa and Alex Tamkin and Miles McCain and Saffron Huang and Esin Durmus and Sarah Heck and Jared Mueller and Jerry Hong and Stuart Ritchie and Tim Belonax and Kevin K. Troy and Dario Amodei and Jared Kaplan and Jack Clark and Deep Ganguli},
howpublished={	arXiv:2503.04761 [cs.CY]},
  year={2025}
}

@article{peng2023impact,
  title={The impact of {AI} on developer productivity: Evidence from {GitHub} {copilot}},
  author={Peng, Sida and Kalliamvakou, Eirini and Cihon, Peter and Demirer, Mert},
  journal={arXiv preprint arXiv:2302.06590},
  year={2023}
}

@article{cui2024effects,
  title={The effects of generative {AI} on high-skilled work: Evidence from three field experiments with software developers},
  author={Cui, Kevin Zheyuan and Demirer, Mert and Jaffe, Sonia and Musolff, Leon and Peng, Sida and Salz, Tobias},
  journal={Management Science},
  year={2026},
  publisher={INFORMS}
}

@article{dillon2025early,
  title={Early Impacts of {M365 Copilot}},
  author={Dillon, Eleanor Wiske and Jaffe, Sonia and Peng, Sida and Cambon, Alexia},
  journal={arXiv preprint arXiv:2504.11443},
  year={2025}
}

@article{brynjolfsson2025generative,
  title={Generative {AI} at work},
  author={Brynjolfsson, Erik and Li, Danielle and Raymond, Lindsey},
  journal={The Quarterly Journal of Economics},
  pages={889--942},
  year={2025},
  publisher={Oxford University Press}
}

@article{noy2023experimental,
  title={Experimental evidence on the productivity effects of generative artificial intelligence},
  author={Noy, Shakked and Zhang, Whitney},
  journal={Science},
  volume={381},
  number={6654},
  pages={187--192},
  year={2023},
  publisher={American Association for the Advancement of Science}
}

@article{bresnahan1995general,
  title={General purpose technologies: ``Engines of growth''?},
  author={Bresnahan, Timothy F and Trajtenberg, Manuel},
  journal={Journal of Econometrics},
  volume={65},
  number={1},
  pages={83--108},
  year={1995},
  publisher={Elsevier}
}

@techreport{microsoftaiei2025,
  title        = {{AI} Diffusion Report: Where {AI} is most used, developed, and built},
  author       = {{Microsoft}},
  year         = {2025},
  institution  = {Microsoft Corporation},
  url          = {https://www.microsoft.com/en-us/research/wp-content/uploads/2025/10/Microsoft-AI-Diffusion-Report.pdf},
  note         = {Accessed: 2025-11-06}
}

@techreport{appel2025anthropic,
  title        = {Anthropic Economic Index report: Uneven geographic and enterprise {AI} adoption},
  author       = {Ruth Appel and Peter McCrory and Alex Tamkin and 
Miles McCain and Tyler Neylon and Michael Stern},
  year         = {2025},
  institution  = {Anthropic},
  url          = {https://www.anthropic.com/research/anthropic-economic-index-september-2025-report},
  note         = {Accessed: 2025-10-05}
}

@online{anthropic2026aeiv4,
        author = {Ruth Appel and Maxim Massenkoff and Peter McCrory and Miles McCain and Ryan Heller and Tyler Neylon and Alex Tamkin},
        title = {Anthropic Economic Index report: economic primitives},
        date = {2026-01-15},
        year = {2026},
        url = {https://www.anthropic.com/research/anthropic-economic-index-january-2026-report},
}

@online{anthropic2026aeiv5,
        author = {Maxim Massenkoff and Eva Lyubich and Peter McCrory and Ruth Appel and Ryan Heller},
        title = {Anthropic Economic Index report: Learning curves},
        date = {2026-03-24},
        year = {2026},
        url = {https://www.anthropic.com/research/economic-index-march-2026-report},
}

@techreport{jaffegenaiworkplaces,
  title        = {Generative {AI} in Real-World Workplaces: The Second Microsoft Report on {AI} and Productivity Research},
  author       = {Jaffe, Sonia and Shah, Neha Parikh and Butler, Jenna and Farach, Alex and Cambon, Alexia and Hecht, Brent and Schwarz, Michael and Teevan, Jaime},
  year         = {2024},
  institution  = {Microsoft Corporation},
  url          = {https://www.microsoft.com/en-us/research/wp-content/uploads/2024/07/Generative-AI-in-Real-World-Workplaces.pdf},
  note         = {Accessed: 2025-10-12}
}

@misc{microsoftaidiffusionmethod,
      title={Measuring {AI} Diffusion: A Population-Normalized Metric for Tracking Global {AI} Usage}, 
      author={Amit Misra and Jane Wang and Scott McCullers and Kevin White and Juan Lavista Ferres},
      year={2025},
      eprint={2511.02781},
      archivePrefix={arXiv},
      primaryClass={cs.CY},
      url={https://arxiv.org/abs/2511.02781}, 
}

@inproceedings{zhaowildchat,
  title={{WildChat: 1M ChatGPT} Interaction Logs in the Wild},
  author={Zhao, Wenting and Ren, Xiang and Hessel, Jack and Cardie, Claire and Choi, Yejin and Deng, Yuntian},
  booktitle={The Twelfth International Conference on Learning Representations},
year={2024}
}

@techreport{acemoglumacroai,
  author      = {Daron Acemoglu},
  title       = {The Simple Macroeconomics of AI},
  institution = {National Bureau of Economic Research},
  type        = {Working Paper},
  number      = {32487},
  year        = {2024},
  month       = {May},
  doi         = {10.3386/w32487},
  url         = {https://www.nber.org/papers/w32487}
}

@article{brynjolfsson2018machines,
Author = {Brynjolfsson, Erik and Mitchell, Tom and Rock, Daniel},
Title = {What can machines learn, and what does it mean for occupations and the economy?},
Journal = {AEA Papers and Proceedings},
Volume = {108},
Year = {2018},
Month = {May},
Pages = {43–47}
}

@article{collisaicontribution,
  author    = {Avinash Collis and Erik Brynjolfsson},
  title     = {{AI}'s Overlooked \$97 Billion Contribution to the Economy},
  journal   = {The Wall Street Journal},
  year      = {2025},
  month     = {August},
  day       = {3},
  url       = {https://www.wsj.com/opinion/ais-overlooked-97-billion-contribution-to-the-economy-users-service-da6e8f55},
  note      = {Accessed: 2025-10-12}
}

@techreport{brynjolfssonagendaai,
  author      = {Erik Brynjolfsson and Anton Korinek and Ajay Agrawal},
  title       = {A Research Agenda for the Economics of Transformative {AI}},
  institution = {National Bureau of Economic Research},
  type        = {Working Paper},
  number      = {34256},
  year        = {2025},
  month       = {September},
  doi         = {10.3386/w34256},
  url         = {https://www.nber.org/papers/w34256}
}

@article{brynjolfsson2017ml,
    author = {Erik Brynjolfsson and Tom Mitchell},
    title = {What can machine learning do? Workforce implications},
    journal = {Science},
    year = {2017},
    month = {December},
    volume = {358},
    number = {6370},
    pages = {1530-1534}
}

@misc{tomlinsonworking,
title={Working with {AI}: Measuring the Applicability of {AI} to Occupations}, 
      author={Kiran Tomlinson and Sonia Jaffe and Will Wang and Scott Counts and Siddharth Suri},
      year={2025},
      howpublished={arXiv:2507.07935 [cs.AI]}
}

@misc{chatterji2025chatgpt,
 title = {How People Use {ChatGPT}},
 author = {Aaron Chatterji and  Tom Cunningham and Christopher Ong and  Carl Shan and David Deming and Z\"{o}e Hitzig and  Kevin Wadman},
 month = {September},
 day = {15}, 
 year = {2025}, 
 hopublished = {NBER Working Paper}
}

@inproceedings{wan2024tnt-llm,
author = {Wan, Mengting and Safavi, Tara and Jauhar, Sujay Kumar and Kim, Yujin and Counts, Scott and Neville, Jennifer and Suri, Siddharth and Shah, Chirag and White, Ryen W. and Yang, Longqi and Andersen, Reid and Buscher, Georg and Joshi, Dhruv and Rangan, Nagu},
title = {TnT-LLM: Text Mining at Scale with Large Language Models},
booktitle = {KDD},
year = {2024},
month = {March},
url = {https://www.microsoft.com/en-us/research/publication/tnt-llm-text-mining-at-scale-with-large-language-models/},
}

\clearpage
\appendix
\section{Figures \& Tables} \label{sec:appendixa}
\setcounter{figure}{0}
\renewcommand{\thefigure}{A\arabic{figure}}
\setcounter{topnumber}{2}
\setcounter{totalnumber}{4}
\renewcommand{\topfraction}{0.95}
\renewcommand{\textfraction}{0.05}
\renewcommand{\floatpagefraction}{0.8}

\begin{figure}[htb]
    \centering
    \includegraphics[width=0.98\linewidth]{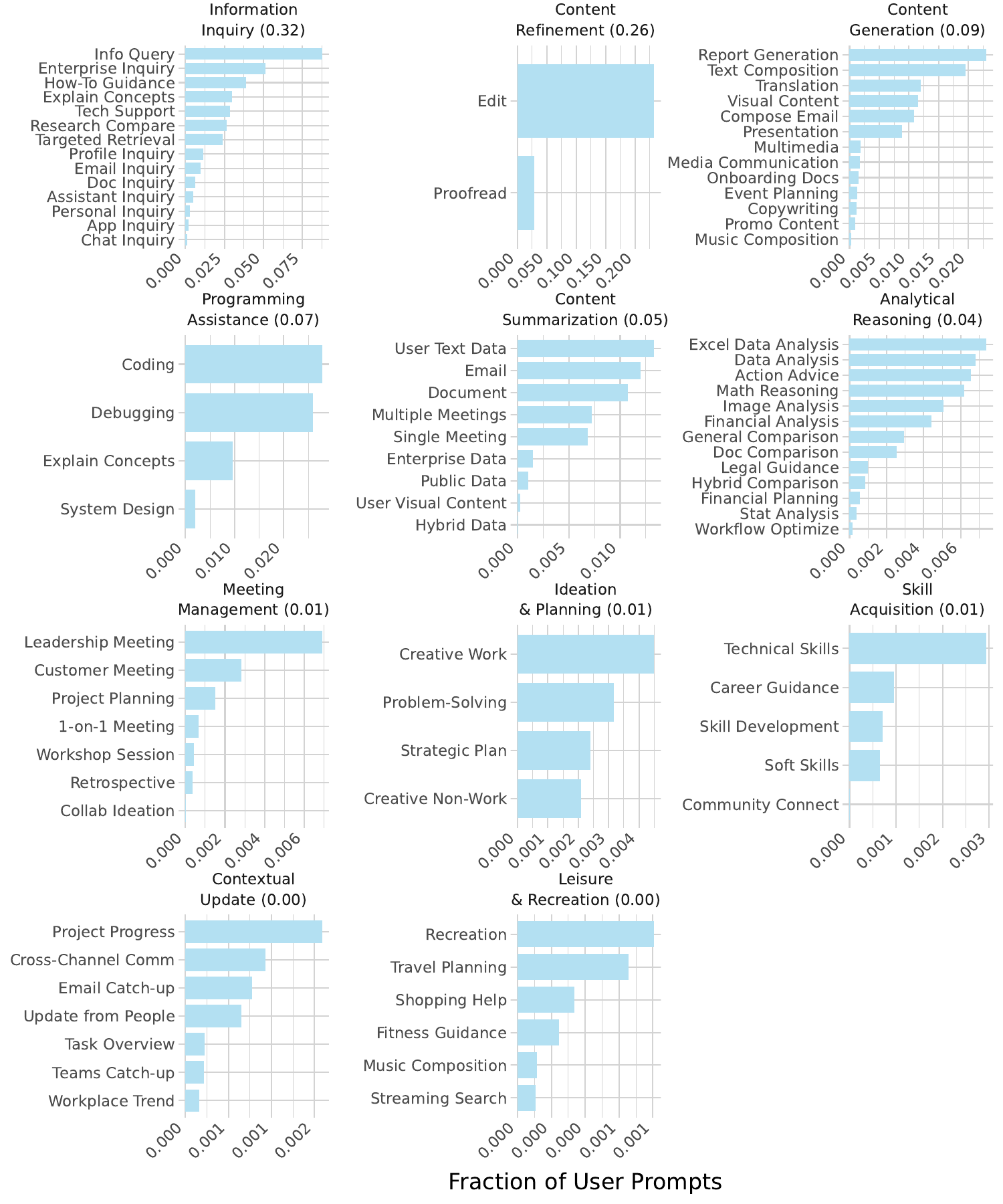}
    \caption{Second-level intents for all top-level intents.}
    \label{fig:intentl2all}
\end{figure}

\begin{figure}[htb]
    \centering
    \includegraphics[width=0.95\linewidth]{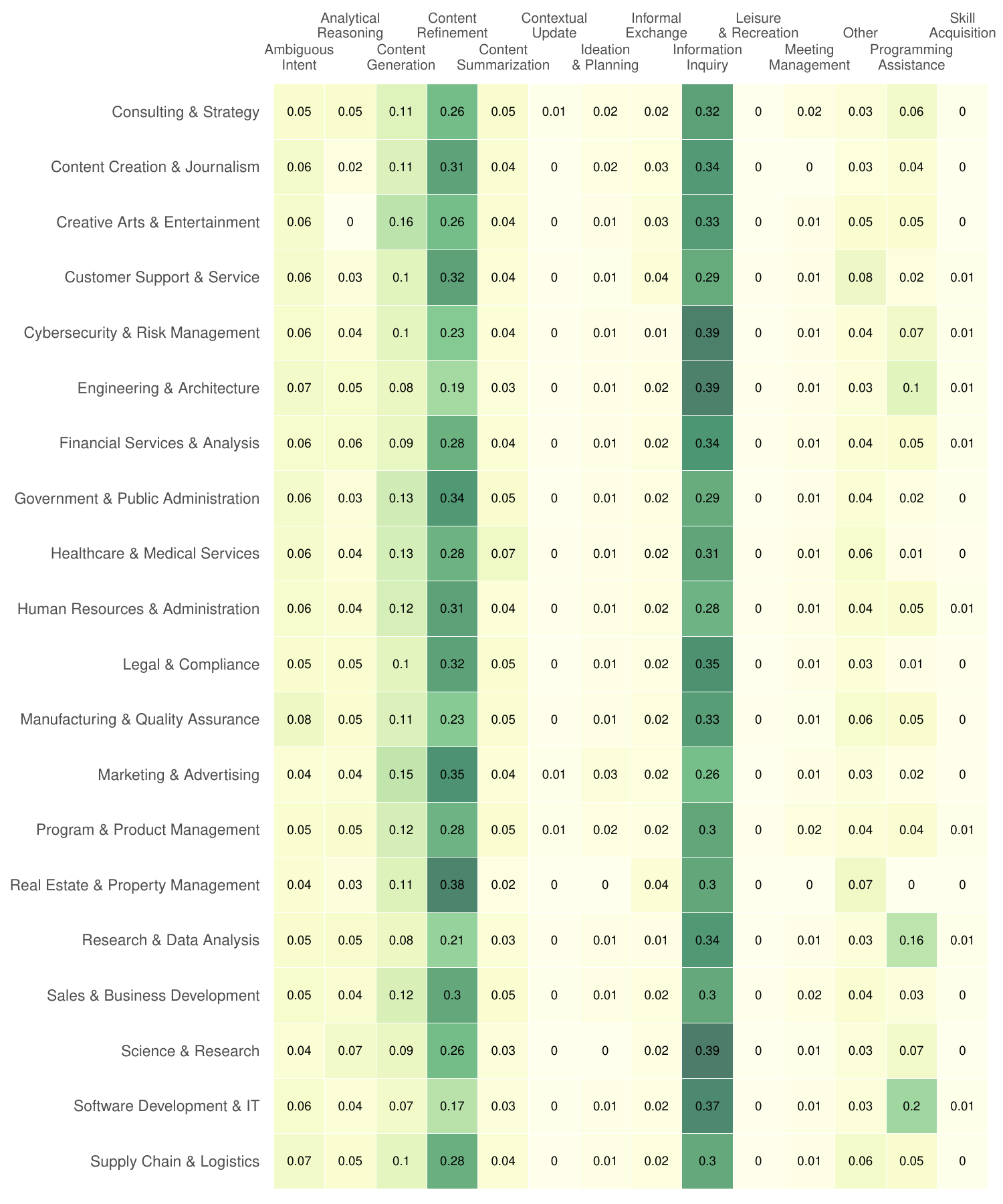}
    \caption{Top-level intent by Occupational Group.}
    \label{fig:intentoccgroupall}
\end{figure}

\begin{figure}[htb]
    \centering
    \includegraphics[width=0.95\linewidth]{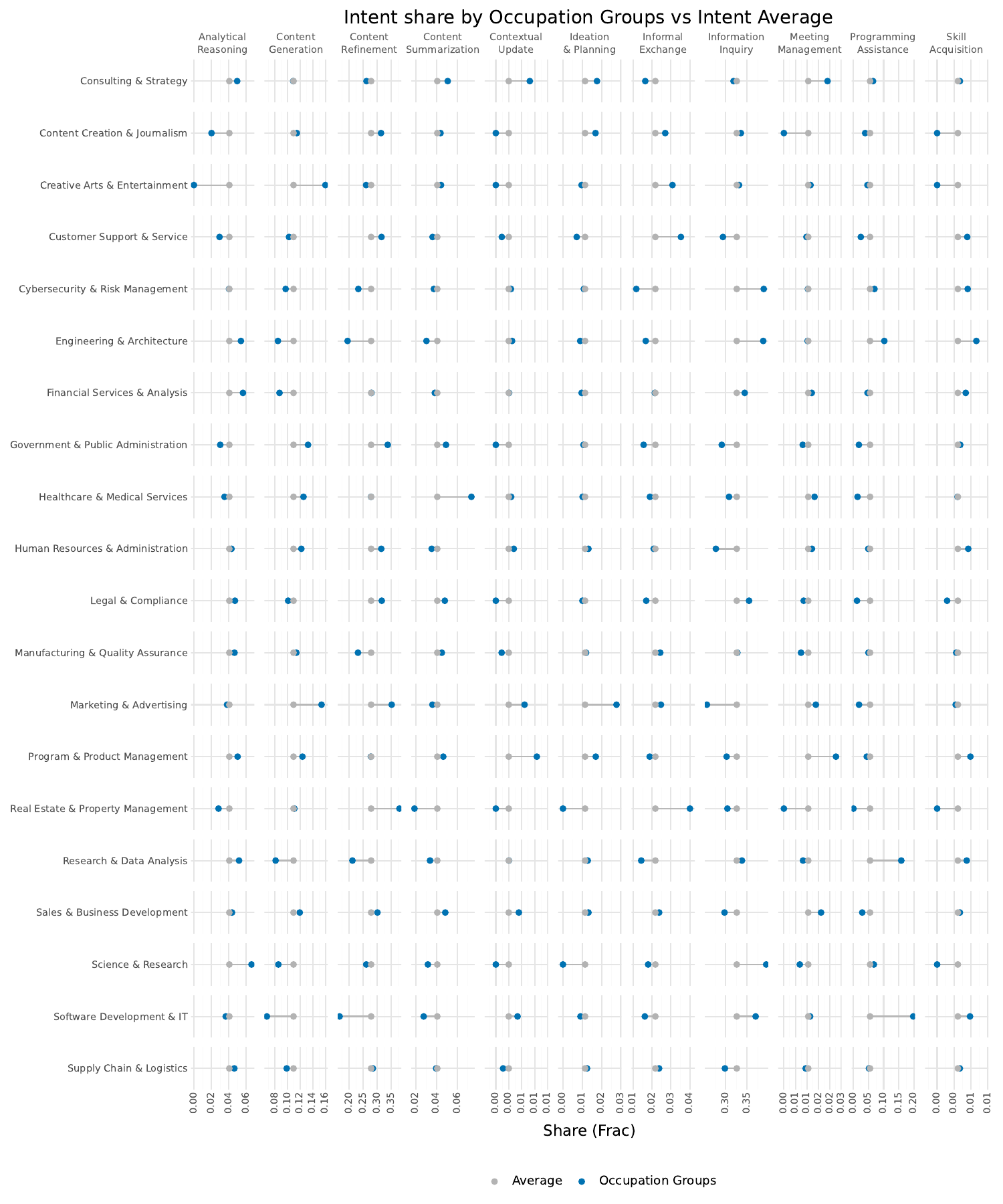}
    \caption{Top-level intent by Occupational Group, showing different from overall mean.}
    \label{fig:intentoccgroup_delta}
\end{figure}

\begin{figure}[h]
    \centering
    \subfigure[IWA skew: User Goal v Labor Market]{
        \includegraphics[width=0.95\linewidth]{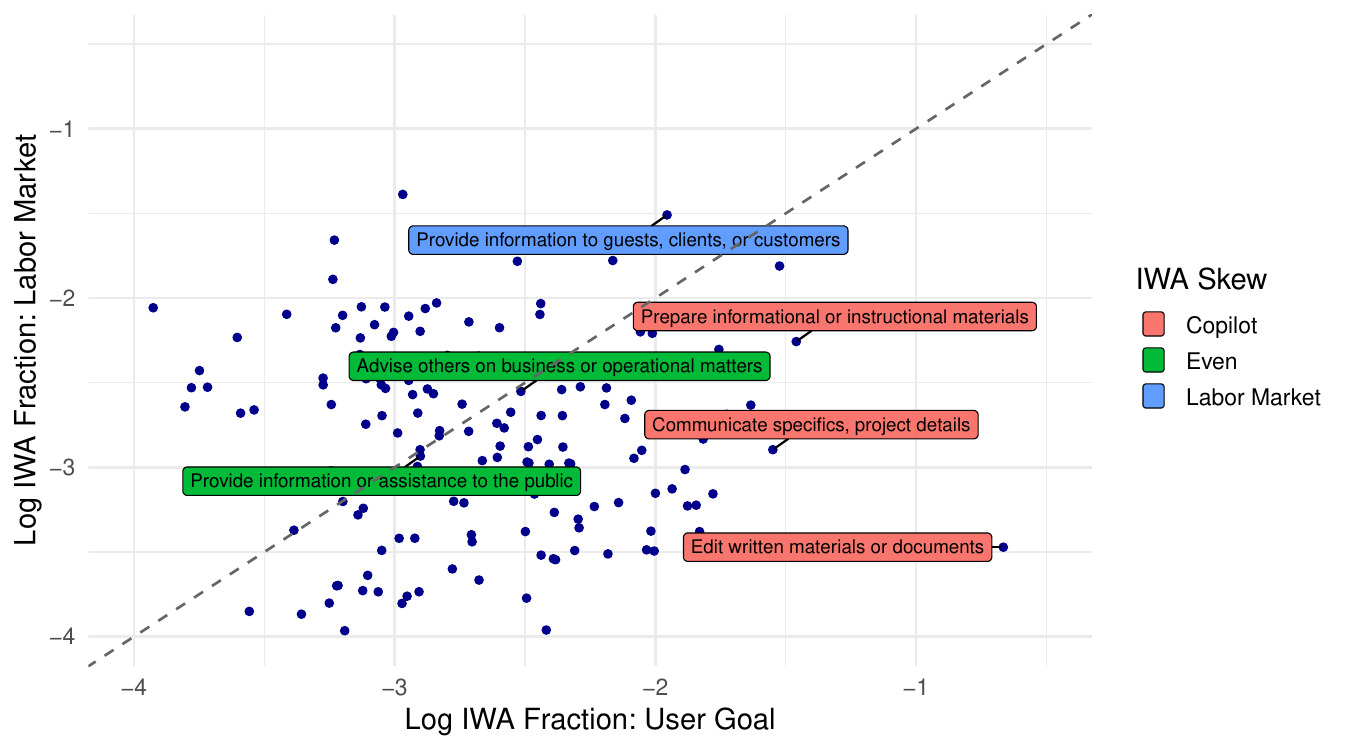}
    }
    \hfill
    \subfigure[IWA skew: Copilot Action v Labor Market]{
        \includegraphics[width=0.95\linewidth]{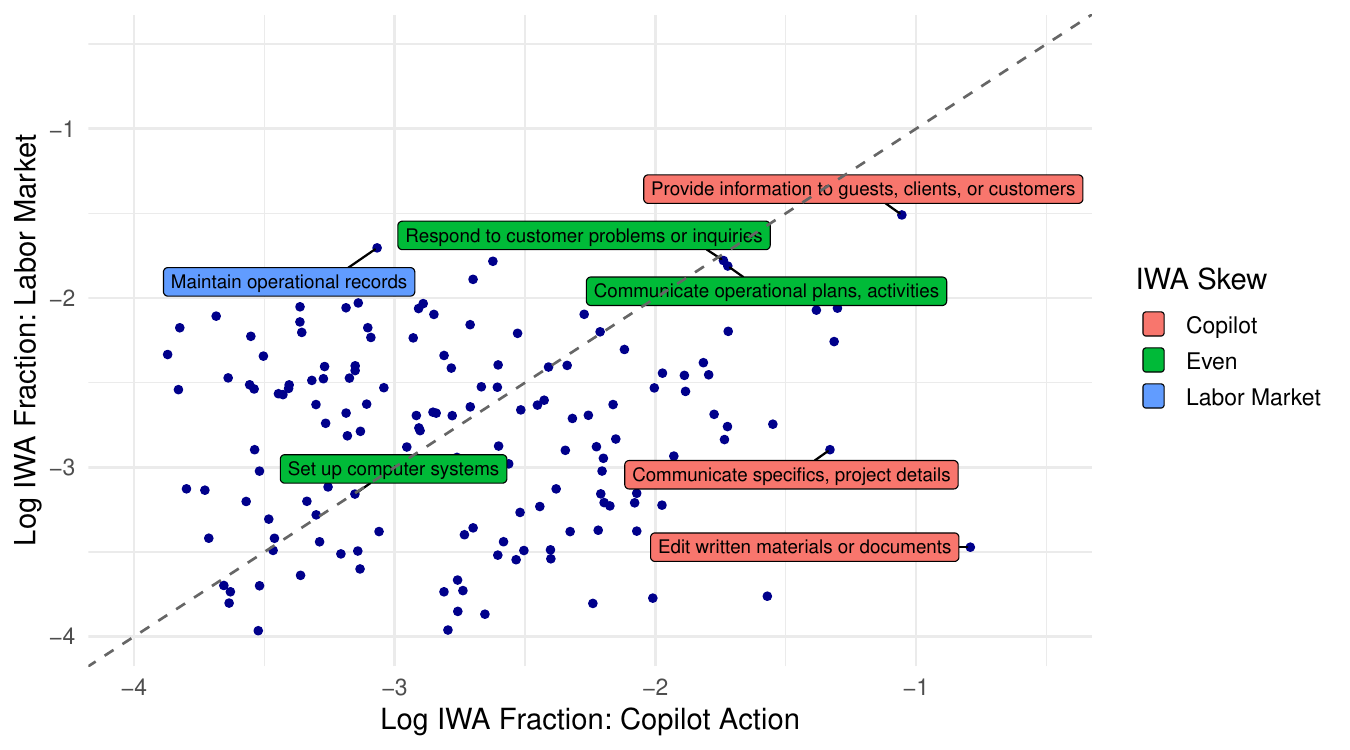}
    }
    \caption{Share of IWAs done in M365 Copilot vs. in the labor market as a whole, user goal (a; $r = -0.01$) and Copilot action (b; $r = 0.17$). Example IWAs shown skew toward Copilot, toward the labor market, or are about even in share. Only IWAs in Copilot data shown.}
    \label{fig:iwa_user_copilot_market}
\end{figure}

\begin{figure}[h]
    \centering
    \includegraphics[width=0.95\linewidth]{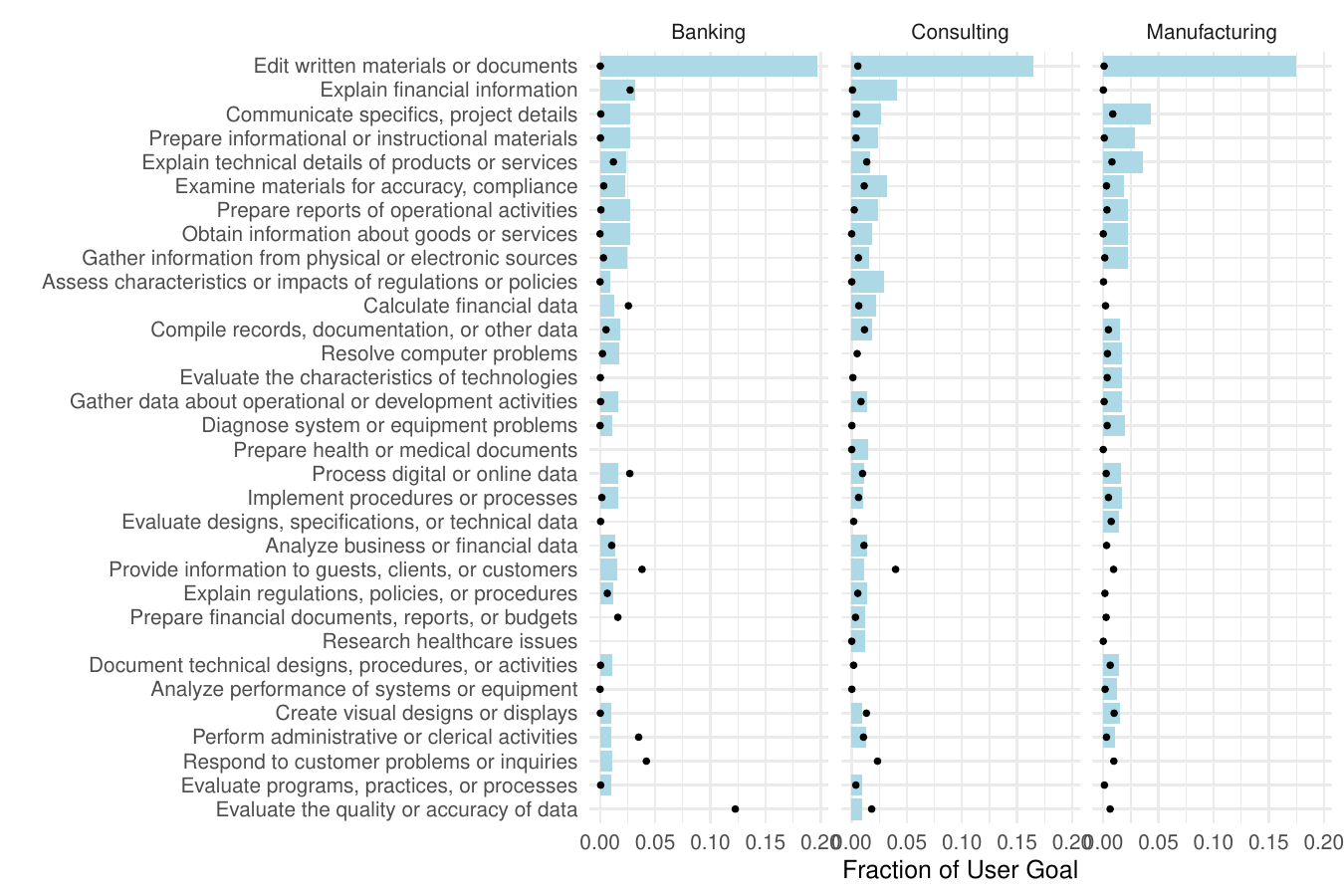}
    \caption{IWAs for each of three example industries. IWAs shown account for 0.9\% or more of activity. Dots shown estimate percent of IWA done within each sector based on Occupational Employment and Wage Statistics data.}
    \label{fig:iwa_industry_labor}
\end{figure}

\begin{figure}[htb]
    \centering
    \includegraphics[width=0.95\linewidth]{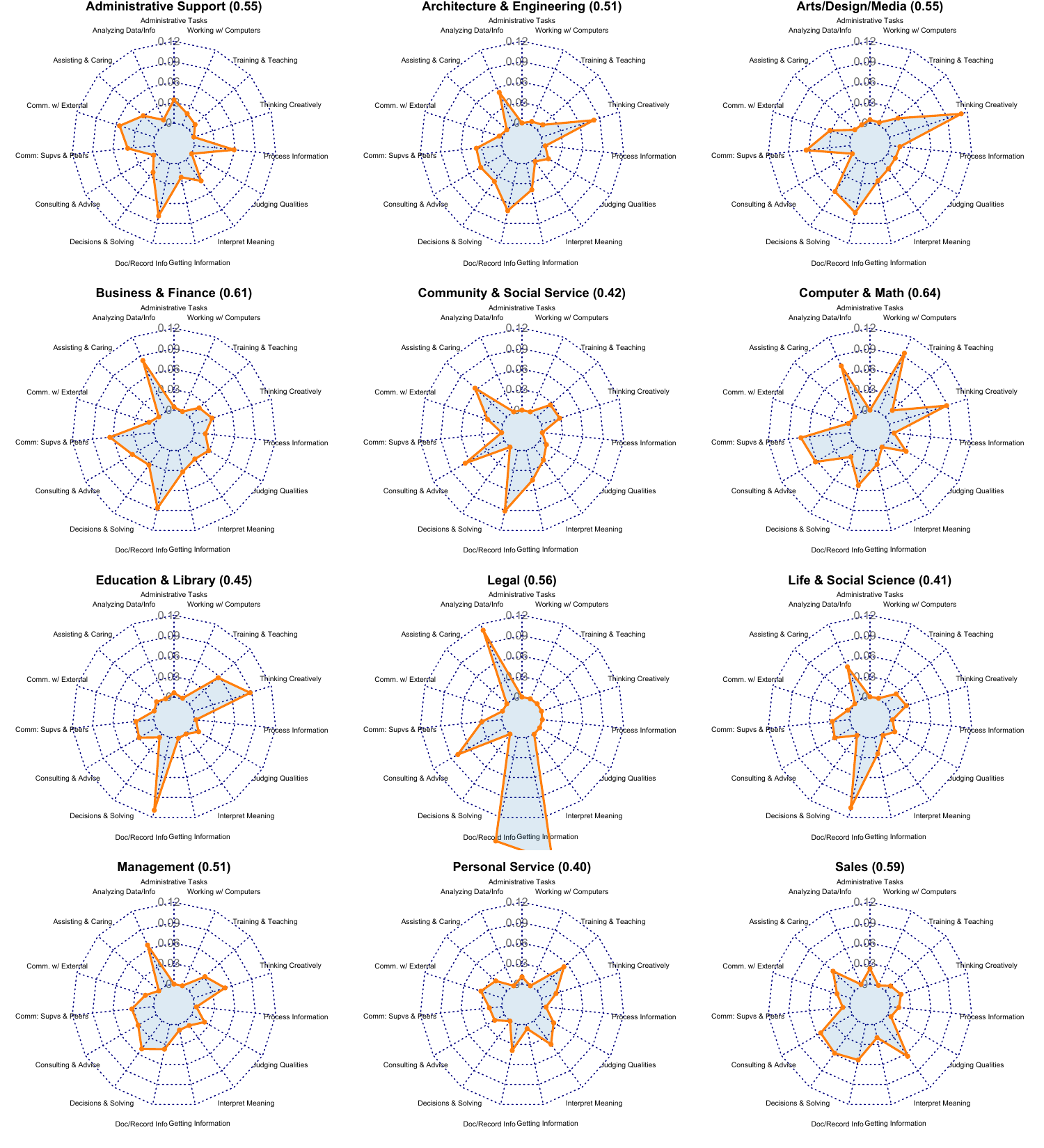}
    \caption{AI Applicability score by job family, broken out into GWAs.}
    \label{fig:job_family_radar_all}
\end{figure}

\newcolumntype{Y}{>{\raggedright\arraybackslash}X}

\begin{table}[!ht]
\centering
\caption{Full list of IWAs in primary work activity dataset.}
\tiny
\setlength{\tabcolsep}{6pt}
\begin{tabular}{p{0.36\textwidth}rr p{0.36\textwidth}rr}
    \toprule
    \cmidrule(lr){1-3} \cmidrule(lr){4-6}
    \textbf{IWA} & \makecell{\textbf{User}\\\textbf{Goal}} & \makecell{\textbf{Copilot}\\\textbf{Action}} &
    \textbf{IWA} & \makecell{\textbf{User}\\\textbf{Goal}} & \makecell{\textbf{Copilot}\\\textbf{Action}} \\
    \midrule
Adjust equipment to ensure adequate performance & 0.00 & 0.00 & Identify business or organizational opportunities & 0.00 & 0.00 \\ 
  Advise others on business or operational matters & 0.00 & 0.01 & Implement procedures or processes & 0.02 & 0.01 \\ 
  Advise others on educational or vocational matters & 0.00 & 0.00 & Implement security measures for computer systems & 0.00 & 0.00 \\ 
  Advise others on financial matters & 0.00 & 0.01 & Inspect conditions of materials or products & 0.00 & 0.00 \\ 
  Advise others on healthcare or wellness issues & 0.00 & 0.00 & Inspect completed work or finished products & 0.00 & 0.00 \\ 
  Advise others on legal or regulatory matters & 0.00 & 0.01 & Inspect facilities or equipment & 0.00 & 0.00 \\ 
  Advise others on products or services & 0.00 & 0.02 & Interpret language, cultural, or religious information & 0.00 & 0.01 \\ 
  Advise others on the design or use of technologies & 0.00 & 0.03 & Interview people to obtain information & 0.00 & 0.00 \\ 
  Advise others on workplace health or safety issues & 0.00 & 0.00 & Investigate incidents or accidents & 0.00 & 0.00 \\ 
  Advise patients or clients on medical issues & 0.00 & 0.00 & Investigate individuals' background or activities & 0.00 & 0.00 \\ 
  Advocate for individual or community needs & 0.00 & 0.00 & Investigate organizational or operational problems & 0.00 & 0.00 \\ 
  Analyze biological or chemical substances or data & 0.00 & 0.00 & Issue documentation & 0.00 & 0.01 \\ 
  Analyze business or financial data & 0.01 & 0.00 & Maintain current knowledge in area of expertise & 0.00 & 0.00 \\ 
  Analyze business or financial risks & 0.00 & 0.00 & Maintain computer, or technical equipment & 0.00 & 0.00 \\ 
  Analyze data to improve operations & 0.01 & 0.00 & Maintain health or medical records & 0.00 & 0.00 \\ 
  Analyze health or medical data & 0.00 & 0.00 & Maintain operational records & 0.00 & 0.00 \\ 
  Analyze market or industry conditions & 0.00 & 0.00 & Maintain safety or security & 0.00 & 0.00 \\ 
  Analyze performance of systems or equipment & 0.01 & 0.00 & Maintain sales or financial records & 0.00 & 0.00 \\ 
  Analyze scientific or applied data & 0.00 & 0.00 & Manage budgets or finances & 0.00 & 0.00 \\ 
  Arrange displays or decorations & 0.00 & 0.00 & Manage control systems or activities & 0.00 & 0.00 \\ 
  Assess regulations or policies & 0.01 & 0.00 & Manage human resources activities & 0.00 & 0.00 \\ 
  Assign work to others & 0.00 & 0.00 & Mark materials or objects for identification & 0.00 & 0.00 \\ 
  Assist individuals with paperwork & 0.00 & 0.03 & Measure materials, products, or equipment & 0.00 & 0.00 \\ 
  Assist others to access services or resources & 0.00 & 0.02 & Monitor equipment operation & 0.00 & 0.00 \\ 
  Assist scientists, with projects or research & 0.00 & 0.00 & Monitor external affairs, trends, or events & 0.00 & 0.00 \\ 
  Calculate financial data & 0.01 & 0.01 & Monitor financial data or activities & 0.00 & 0.00 \\ 
  Coach others & 0.00 & 0.00 & Monitor individual behavior or performance & 0.00 & 0.00 \\ 
  Collect data about consumer needs or opinions & 0.00 & 0.00 & Monitor operation of computer technologies & 0.00 & 0.00 \\ 
  Collect information about patients or clients & 0.00 & 0.00 & Monitor operations to ensure adequate performance & 0.00 & 0.00 \\ 
  Communicate sustainability information & 0.00 & 0.00 & Monitor operations for compliance with regulations & 0.00 & 0.00 \\ 
  Communicate with others about business strategies & 0.00 & 0.01 & Monitor resources or inventories & 0.00 & 0.00 \\ 
  Communicate operational plans, activities & 0.03 & 0.02 & Notify others of emergencies or problems & 0.01 & 0.00 \\ 
  Communicate specifics, project details & 0.03 & 0.05 & Obtain formal documentation or authorization & 0.01 & 0.00 \\ 
  Compile records, documentation, or other data & 0.02 & 0.02 & Obtain information about goods or services & 0.02 & 0.00 \\ 
  Confer with clients to determine needs & 0.00 & 0.01 & Operate computer systems or equipment & 0.00 & 0.00 \\ 
  Consult legal materials or public records & 0.01 & 0.00 & Perform administrative or clerical activities & 0.01 & 0.00 \\ 
  Coordinate activities with clients or agencies & 0.00 & 0.00 & Perform human resources activities & 0.00 & 0.00 \\ 
  Coordinate group, community, or public activities & 0.00 & 0.00 & Perform recruiting or hiring activities & 0.00 & 0.00 \\ 
  Coordinate regulatory compliance activities & 0.00 & 0.00 & Plan events or programs & 0.00 & 0.00 \\ 
  Coordinate with others to resolve problems & 0.00 & 0.00 & Plan work activities & 0.01 & 0.00 \\ 
  Counsel others about personal matters & 0.00 & 0.00 & Prepare documentation for contracts or permits & 0.00 & 0.00 \\ 
  Create artistic designs or performances & 0.00 & 0.00 & Prepare financial documents, reports, or budgets & 0.00 & 0.00 \\ 
  Create visual designs or displays & 0.02 & 0.01 & Prepare health or medical documents & 0.01 & 0.00 \\ 
  Design computer systems or applications & 0.00 & 0.00 & Prepare informational or instructional materials & 0.03 & 0.05 \\ 
  Design materials or devices & 0.00 & 0.00 & Prepare legal or regulatory documents & 0.00 & 0.00 \\ 
  Determine operational methods or procedures & 0.01 & 0.01 & Prepare proposals or grant applications & 0.00 & 0.00 \\ 
  Determine resource needs of projects or operations & 0.00 & 0.00 & Prepare reports of operational activities & 0.02 & 0.02 \\ 
  Determine values or prices of goods or services & 0.00 & 0.00 & Prepare schedules for services or facilities & 0.00 & 0.00 \\ 
  Develop business or marketing plans & 0.00 & 0.00 & Present research or technical information & 0.00 & 0.02 \\ 
  Develop educational programs, plans, or procedures & 0.00 & 0.00 & Process digital or online data & 0.02 & 0.01 \\ 
  Develop financial or business plans & 0.00 & 0.00 & Program computer systems or production equipment & 0.00 & 0.00 \\ 
  Develop marketing or promotional materials & 0.01 & 0.01 & Promote products, services, or programs & 0.00 & 0.00 \\ 
  Develop models of systems, processes, or products & 0.00 & 0.00 & Provide general assistance to others & 0.00 & 0.00 \\ 
  Develop news, entertainment, or artistic content & 0.00 & 0.00 & Provide information or assistance to the public & 0.00 & 0.01 \\ 
  Develop operational procedures or standards & 0.01 & 0.01 & Provide information to guests, clients, or customers & 0.01 & 0.09 \\ 
  Develop organizational goals or objectives & 0.00 & 0.00 & Provide support or encouragement to others & 0.00 & 0.00 \\ 
  Develop organizational policies, systems, or processes & 0.00 & 0.00 & Purchase goods or services & 0.00 & 0.00 \\ 
  Develop patient or client care or treatment plans & 0.00 & 0.00 & Read materials to inform work processes & 0.03 & 0.05 \\ 
  Develop professional relationships or networks & 0.00 & 0.00 & Reconcile financial data & 0.00 & 0.00 \\ 
  Develop recipes or menus & 0.00 & 0.00 & Research healthcare issues & 0.01 & 0.00 \\ 
  Develop safety standards, policies, or procedures & 0.00 & 0.00 & Research historical or social issues & 0.00 & 0.00 \\ 
  Develop technical specifications & 0.00 & 0.00 & Research laws, precedents, or other legal data & 0.00 & 0.00 \\ 
  Diagnose system or equipment problems & 0.01 & 0.01 & Research organizational behavior or performance & 0.00 & 0.00 \\ 
  Discuss legal matters & 0.00 & 0.00 & Research technology designs or applications & 0.01 & 0.00 \\ 
  Document technical designs, procedures, or activities & 0.01 & 0.01 & Resolve computer problems & 0.02 & 0.01 \\ 
  Edit written materials or documents & 0.22 & 0.16 & Resolve personnel or operational problems & 0.00 & 0.00 \\ 
  Estimate project development or operational costs & 0.00 & 0.00 & Respond to customer problems or inquiries & 0.01 & 0.02 \\ 
  Evaluate condition of financial assets, property & 0.00 & 0.00 & Schedule appointments & 0.00 & 0.00 \\ 
  Evaluate designs, specifications, or technical data & 0.01 & 0.01 & Schedule operational activities & 0.00 & 0.00 \\ 
  Evaluate patient condition or treatment options & 0.00 & 0.00 & Select materials for operations or projects & 0.00 & 0.00 \\ 
  Evaluate personnel capabilities or performance & 0.00 & 0.00 & Set up computer systems & 0.00 & 0.00 \\ 
  Evaluate production inputs or outputs & 0.00 & 0.00 & Set up equipment & 0.00 & 0.00 \\ 
  Evaluate programs, practices, or processes & 0.01 & 0.01 & Signal others to coordinate work activities & 0.00 & 0.00 \\ 
  Evaluate project feasibility & 0.00 & 0.00 & Study details of artistic productions & 0.00 & 0.00 \\ 
  Evaluate the characteristics of technologies & 0.01 & 0.00 & Teach academic or vocational subjects & 0.00 & 0.00 \\ 
  Evaluate the quality or accuracy of data & 0.01 & 0.01 & Teach life skills & 0.00 & 0.00 \\ 
  Examine financial activities, operations, or systems & 0.00 & 0.00 & Teach safety procedures or standards to others & 0.00 & 0.00 \\ 
  Examine materials for accuracy, compliance & 0.02 & 0.01 & Test performance of computer or information systems & 0.00 & 0.00 \\ 
  Execute financial transactions & 0.00 & 0.00 & Test performance of equipment or systems & 0.00 & 0.00 \\ 
  Explain financial information & 0.01 & 0.01 & Train others on health or medical topics & 0.00 & 0.00 \\ 
  Explain medical information to patients or family & 0.00 & 0.01 & Train others on operational or work procedures & 0.00 & 0.00 \\ 
  Explain regulations, policies, or procedures & 0.01 & 0.02 & Train others to use equipment or products & 0.00 & 0.00 \\ 
  Explain technical details of products or services & 0.02 & 0.04 & Verify personal information & 0.00 & 0.00 \\ 
  Gather data about operational activities & 0.01 & 0.00 & Write material for artistic or commercial purposes & 0.01 & 0.01 \\ 
  Gather information from physical or electronic source & 0.02 & 0.00 &  &  &  \\ 
    \bottomrule
\end{tabular}
\label{tab:iwa_full_list}
\end{table}

\clearpage
\section{Classifiers}

\begin{tcolorbox}[title=Work Classifier, fonttitle=\bfseries]
\small
You are an AI assistant tasked with classifying a sequence of users' Conversations to AI Chat(Copilot) as either **Personal** or **Work** related or **unclear**.
\\

**Categories**:\\ 
- **Personal** – The user’s prompt is about personal or everyday matters (e.g. family, health, home, hobbies, or general life). \\ 
- **Work** – The user’s prompt is about business, work, or professional matters (e.g. office tasks, projects, corporate information, or job-related topics).\\
- **Unclear** – The user’s prompt is not about personal or Work, or it is not possible to classify the prompts to any category\\

**Instructions**: Read the sequence of conversations (upto past 3 conversations if available is given to you) and determine which category the last conversation (also labelled as "conversation to be classified") falls into. **Respond with only the single word** “Personal” **or** “Work” **or** “Unclear” accordingly, and nothing else. Do not add any explanations or extra words.
\\
**Examples**:\\  
- **User Prompt:** Turn1: "I need to buy groceries and pick up my kids from school."  
Turn2: "Recommend good restaurant on the way back?"
  *Your Response:* Personal  
\\
- **User Prompt:** Turn1: "Prepare the slides for Monday’s sales meeting."  
Turn2: "Modify the slide to include a summary slide for executives."
  **Your Response:** Work\\
\\
Now, classify the **Conversation**.\\

\{\{User Prompt\}\}\\

**Your Response:**

\end{tcolorbox}

\end{document}